\documentclass[11pt,twocolumn]{article}
\usepackage{times}
\usepackage{balance}
\usepackage{authblk}
\usepackage{hyperref}
\usepackage{graphicx}
\usepackage[a4paper,margin=0.75in]{geometry} 

\usepackage{booktabs}
\usepackage{titlesec}
\titlespacing*{\section}{0pt}{1ex}{0.5ex}
\titlespacing*{\subsection}{0pt}{0.8ex}{0.4ex}

\title{Towards Scalable Federated Container Orchestration: The CODECO Approach}

\author[1]{Rute C. Sofia\,}
\author[2]{Josh Salomon}
\author[2]{Ray Carrol}
\author[3]{Luis Garcés-Erice\,}
\author[3]{Peter Urbanetz}
\author[4]{J{\"u}rgen Gesswein}
\author[5]{Rizkallah Touma}
\author[5]{Alejandro Espinosa}
\author[6]{Luis M. Contreras}
\author[7]{Vasileios Theodorou}
\author[7]{George Papathanail}
\author[8]{Georgios Koukis}
\author[8]{Vassilis Tsaoussidis}
\author[10]{Alberto del Rio}
\author[10]{David Jimenez}
\author[12]{Efterpi Paraskevoulakou}
\author[12]{Panagiotis Karamolegkos}
\author[15]{John Soldatos}
\author[16]{Borja Dorado Nogales}
\author[16]{Alejandro Tjaarda}

\affil[1]{fortiss GmbH, Germany}
\affil[2]{Red Hat, Israel}
\affil[3]{IBM Research Europe - Zurich, Switzerland}
\affil[4]{Siemens AG, Germany}
\affil[5]{i2CAT Foundation, Spain}
\affil[6]{Telef\'onica, Spain}
\affil[7]{Intracom Telecom, Greece}
\affil[8]{Democritus University of Thrace / Athena Research Center, Greece}
\affil[10]{Universidad Polit\'ecnica de Madrid, Spain}
\affil[12]{University of Piraeus Research Centre, Greece}
\affil[15]{Netcompany, Luxembourg}
\affil[16]{Universidad Carlos III de Madrid, Spain}

\begin{document}
\maketitle

\begin{abstract}
This paper presents CODECO, a federated orchestration framework for Kubernetes that addresses the limitations of cloud-centric deployment. CODECO adopts a data–compute–network co-orchestration approach to support heterogeneous infrastructures, mobility, and multi-provider operation.

CODECO extends Kubernetes with semantic application models, partition-based federation, and AI-assisted decision support, enabling context-aware placement and adaptive management of applications and their micro-services across federated environments. A hybrid governance model combines centralized policy enforcement with decentralized execution and learning to preserve global coherence while supporting far Edge autonomy. The paper describes the architecture and core components of CODECO, outlines representative orchestration workflows, and introduces a software-based experimentation framework for reproducible evaluation in federated Edge-Cloud infrastructure environments.
\end{abstract}
\newenvironment{keywords}{\paragraph{Keywords.}}{}


\maketitle


\section{Introduction}
\label{sec:introduction}

The \textit{Cloud–Edge–IoT (CEI)} continuum is reshaping how computation and data processing are deployed, pushing functionality closer to users and devices to meet stringent requirements for low latency, resilience, sustainability, and contextual adaptation. Emerging applications such as smart mobility, industrial automation, and immersive media increasingly rely on distributed execution across heterogeneous, and often intermittently connected, environments.

Despite this evolution, most container orchestration technologies remain largely cloud-centric. Kubernetes (K8s)\footnote{\url{https://kubernetes.io/}}, although widely adopted, assumes relatively stable clusters, centralized control, and a homogeneous infrastructure under a single administrative domain. These assumptions become fragile in far-edge and multi-provider settings, where connectivity is variable, resources are constrained, and governance boundaries differ.

Three limitations are particularly relevant in CEI environments. First, existing tools exhibit limited cross-layer context awareness, with orchestration decisions primarily driven by compute-centric metrics while under-accounting for network conditions and data constraints. Second, prevailing orchestrators are built around centralized control assumptions that do not hold under mobility, intermittent connectivity, and autonomous operation at the far edge. Third, they lack federated abstractions for multi-provider collaboration and controlled information exchange.

\textit{CODECO (COgnitive Decentralized Edge–Cloud Orchestration)}\footnote{\url{https://gitlab.eclipse.org/eclipse-research-labs/codeco-project}} addresses these gaps by augmenting Kubernetes with federated orchestration capabilities, semantic application abstractions, and AI-assisted decision support, as detailed in Sections~\ref{sec:codeco} and~\ref{sec:codeco-fc-arch}. CODECO adopts a data–compute–network co-orchestration paradigm to enable context-aware placement and adaptive management of microservices across distributed CEI infrastructures.

This paper presents a system and architectural perspective on federated Kubernetes orchestration based on the CODECO \textit{Federated Cluster (FC)} framework.Its main contributions are: (i) the design and architecture of a bounded, neighborhood-based federated orchestration framework for the Cloud–Edge–IoT continuum; (ii) a detailed description of the operational workflows combining application intent, cross-layer observability, and AI-assisted decision support; and (iii) an open-source experimentation framework enabling reproducible deployment and evaluation of federated CEI orchestration.

The remainder of this paper is organized as follows.  
Section~\ref{sec:codeco} introduces the design principles and scope of CODECO, including its data–compute–network co-orchestration model and federation and governance approach.  
Section~\ref{sub:context-awareness} discusses how CODECO incorporates context-awareness and energy-aware (greenness) objectives into orchestration decisions across the CEI continuum.  
Section~\ref{sec:codeco-fc-arch} details the federated architecture of CODECO, describing its core components, management layers and timescales, application neighborhoods, and semantic application abstractions.  
Section~\ref{Sec: components} focuses on the operational roles and runtime interactions of the CODECO components, including learning-assisted decision support, network management, scheduling, and metadata handling.  
Section~\ref{sec: workflow} presents representative workflows covering platform installation, federated application deployment, and runtime adaptation.  
Section~\ref{sec:experimentation} introduces the CODECO experimentation strategy and the CODEF software-defined testbed used for reproducible evaluation.  
Section~\ref{sec:future} discusses open research challenges and future directions for federated orchestration across the CEI continuum.  
Section~\ref{sec:relatedwork} surveys related work and provides a comparative analysis with existing multi-cluster and federated orchestration frameworks.  
Finally, Section~\ref{sec:conclusions} concludes the paper and summarizes the main contributions.

\section{CODECO Design Principles and Scope}
\label{sec:codeco}
This section introduces the foundational design principles that guide the architecture and operation of CODECO. It focuses on the conceptual scope of the framework and the system assumptions that shape its orchestration logic. In particular, CODECO departs from cloud-centric orchestration by (i) adopting a data–compute–network co-orchestration paradigm that treats networking, data observability, and computation as first-class scheduling inputs, and (ii) defining a federated governance model that balances centralized policy control with decentralized execution and learning at the far Edge. These principles provide the conceptual basis for the architectural realization described in Section~\ref{sec:codeco-fc-arch}. The remainder of this section elaborates on the co-orchestration model underpinning CODECO and its layered federation, autonomy, and governance approach.

\subsection{Co-Orchestration Across Data, Compute, and Network}
\label{subsec:concept}
At the core of CODECO lies the principle that effective orchestration in CEI environments requires a holistic view of infrastructure and applications. CODECO therefore models the system from three complementary perspectives: networking, data observability, and computation. Orchestration decisions are made by jointly considering these perspectives, enabling placement and adaptation strategies that reflect end-to-end application requirements rather than isolated resource metrics.

\begin{figure*}[t!]
  \centering
  \includegraphics[width=\textwidth]{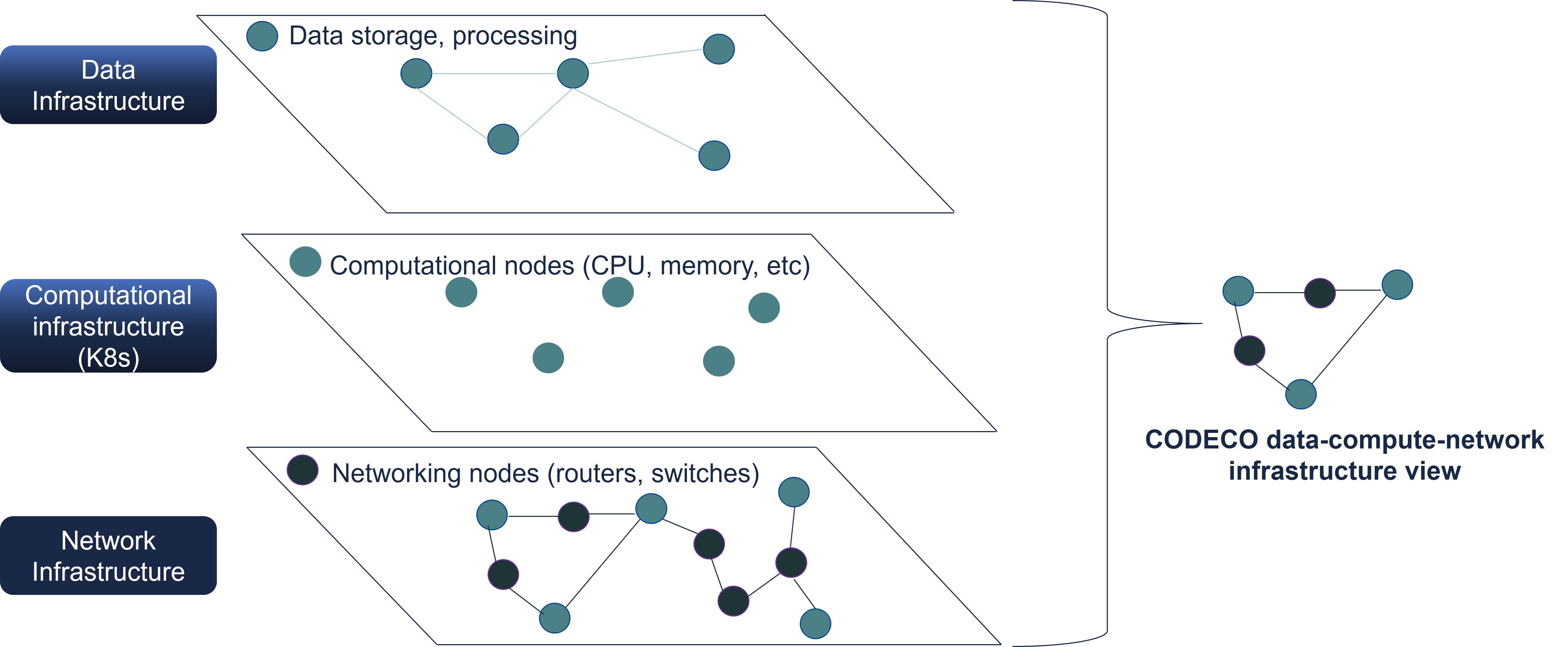}
  \caption{The CODECO data-compute-network approach for deployment of applications and their micro-services across CEI. By observing the data workflow, the network and the computational resources, CODECO provides a finer-grained perspective on the IoT-Edge-Cloud infrastructure, thus providing a better support for the orchestration of applications across Edge-Cloud.}
  \label{fig:codeco-data-compute-network}
\end{figure*}

Figure~\ref{fig:codeco-data-compute-network} illustrates this integrated approach, where application dataflows, network paths, and compute resources are observed and evaluated together to support informed orchestration decisions across the CEI continuum.

First, the \textit{networking perspective} captures both underlay and overlay
connectivity required to deploy and interconnect containerized applications
across Edge-Cloud environments. This includes not only classical QoS indicators
such as latency and bandwidth, but also properties related to mobility and
dynamic topology changes.

Second, the \textit{data observability perspective} captures data dependencies,
locality constraints, and dataflow characteristics associated with microservices.
These properties are essential during initial deployment as well as during
redeployment and adaptation, particularly for data-intensive and latency-sensitive
applications.

Third, the \textit{computational perspective} characterizes the capabilities and
constraints of available nodes, including processing capacity, memory, and
energy-related properties, in order to identify suitable execution environments
for application components.

This integrated perspective \cite{sofia2024framework} reflects the increasing mobility, heterogeneity, and data intensity of CEI applications, where placement decisions must jointly account for computation, networking, and data constraints rather than isolated resource metrics.

\subsection{Federation, Autonomy, Governance Model}
\label{subsec:federation-model}
CODECO adopts a layered governance model tailored to federated CEI environments, where infrastructure spans multiple administrative domains and connectivity conditions vary over time. Rather than relying on fully centralized or fully decentralized control, CODECO defines a hybrid management model that separates governance, coordination, and execution responsibilities. This separation enables global policy consistency while preserving local autonomy and resilience at the far Edge.

The governance layer defines the authoritative desired state of the system. It is responsible for expressing application intent, enforcing organizational and regulatory policies, and maintaining auditability across the federation. In CODECO, governance is centralized at the Hub cluster and leverages Kubernetes-native federation mechanisms provided by \textit{Open Cluster Management (OCM)}\footnote{https://open-cluster-management.io/}, to establish a shared control and governance substrate across clusters. OCM provides well-defined mechanisms for cluster registration, identity management, policy propagation, and application lifecycle coordination using a \textbf{hub-and-spoke }architecture. These capabilities are essential for enforcing organizational policies, security constraints, and compliance requirements across multi-provider environments. The implications of intermittent connectivity and far-edge autonomy are reflected in the architectural design described in Section~\ref{sec:codeco-fc-arch}.

The coordination layer limits the scope of orchestration and information exchange in order to ensure scalability and efficiency. CODECO introduces application \textbf{neighborhoods} as bounded coordination domains that group a subset of Managed Clusters relevant to a specific application. Neighborhoods define where monitoring data is exchanged, where learning components collaborate, and where scheduling decisions are optimized. By constraining coordination to application-specific subsets of clusters, CODECO avoids global state dissemination while still enabling cooperative behaviour across distributed environments. 

The execution layer resides within Managed Clusters and is responsible for enforcing placement, migration, and network configuration decisions using locally available context. This layer operates autonomously and continues to function even when connectivity to the Hub is degraded or temporarily unavailable. Scheduling, adaptation, and learning components at this layer consume centrally defined intent and policies but rely on fresh local observations to make timely decisions. This design ensures resilience and responsiveness at the far edge while preserving convergence toward the desired global state once connectivity is restored.

\section{Context-awareness and Energy-aware Management}
\label{sub:context-awareness}

In existing container orchestration solutions, replication and scaling decisions are typically driven by a limited set of triggers, such as system-level metrics (e.g., CPU utilization or energy consumption thresholds), application-level constraints, or static network policies. While effective in relatively stable Cloud environments, these mechanisms are insufficient for highly dynamic Edge-Cloud settings, where resource availability, connectivity, and execution context may change rapidly.

To address these challenges, CODECO integrates \textit{context-awareness} into the offloading and scheduling process. Context-awareness refers to the capability of a system to exploit knowledge about its operating environment in order to adapt its behavior and perform appropriate actions~\cite{brader1997a,abowd-a}. Context information spans multiple dimensions, including computing context, user context,
and environmental context.

A pre-requisite for context-aware orchestration is the identification of relevant context categories, the definition of suitable models to represent them, and the integration of these models into the orchestration workflow. CODECO considers the following categories of context-aware metrics~\cite{CODECO-D11}:

\begin{itemize}
    \item \textbf{Computational metrics}, such as CPU and memory utilization,
    collected from Kubernetes environments via Prometheus and exposed through ACM.
    \item \textbf{Network metrics}, capturing functional QoS properties (e.g., latency, bandwidth) as well as non-functional aspects such as energy expenditure.
    \item \textbf{Application metrics}, describing application-level QoS and non-functional requirements, including latency bounds, sensitivity levels, certification, and compliance constraints.
    \item \textbf{Data observability metrics}, characterizing data properties such     as freshness, age of information, locality, and compliance constraints.
    \item \textbf{User behavior metrics} reflecting user-related context relevant to far Edge operation, including user location, mobility, preferences, and social or physical proximity~\cite{liu2011a,sofia2023role}.
\end{itemize}

These context dimensions are exposed through CODECO’s observability pipeline and applied in different stages of the application lifetime, as discussed in Section~\ref{subsec:impact-orchestration}.

\subsection{Energy-awareness and Greenness Objectives}
Energy-awareness in CODECO is expressed through the notion of \textbf{greenness}, which captures sustainability objectives at a node and cluster level, such as reduced energy consumption and lower CO\textsubscript{2} footprint. Rather than prescribing a single definition, CODECO allows greenness to be defined through customizable metrics that can be combined with other performance objectives, including latency, resilience, and data locality.

Greenness metrics are treated as optimization criteria rather than hard constraints, enabling trade-offs to be made dynamically based on application intent and operational context.

In the current implementation\footnote{refer to the Eclipse CODECO GitLab https://gitlab.eclipse.org/eclipse-research-labs/codeco-project}, greenness is quantified using the following energy-related metrics:

\begin{itemize}
    \item \textbf{Node Energy ($n_e(i)$)}: energy consumed by node $i$ due to all
    active processes.
    \item \textbf{Link Energy ($l_e(i,j)$)}: energy consumed by node $i$ for
    transmitting data across egress link $j$.
    \item \textbf{Network Energy ($L_e(i)$)}: total energy consumed across all
    egress links of node $i$, defined as $L_e(i)=\sum_{j} l_e(i,j)$.
    \item \textbf{Flow Energy ($F_e(i)$)}: energy associated with application data
    flows originating from node $i$.
\end{itemize}

CODECO facilitates the definition and integration of additional, user-defined greenness metrics, as discussed in Section~\ref{sec:codeco-fc-arch}.

\subsection{Impact on Orchestration Decisions}
\label{subsec:impact-orchestration}

Context-aware and energy-aware metrics directly influence CODECO’s management actions at multiple stages of the application lifecycle:

\begin{itemize}
    \item \textbf{Initial deployment}, where node and cluster selection accounts for current resource availability, network conditions, and energy profiles.
    \item \textbf{Runtime adaptation}, where changes in context (e.g., mobility, congestion, or energy depletion) may trigger workload migration, reconfiguration, or rescheduling.
    \item \textbf{neighborhood re-evaluation}, where sustained performance degradation or operational instability can lead to reassessment of the coordination scope associated with an application.
    \item \textbf{Policy-compliant optimization}, where trade-offs between latency, resilience, and greenness are resolved according to application intent and governance con
\end{itemize}


\begin{figure*}[t!]
  \centering
  \includegraphics[width=\textwidth]{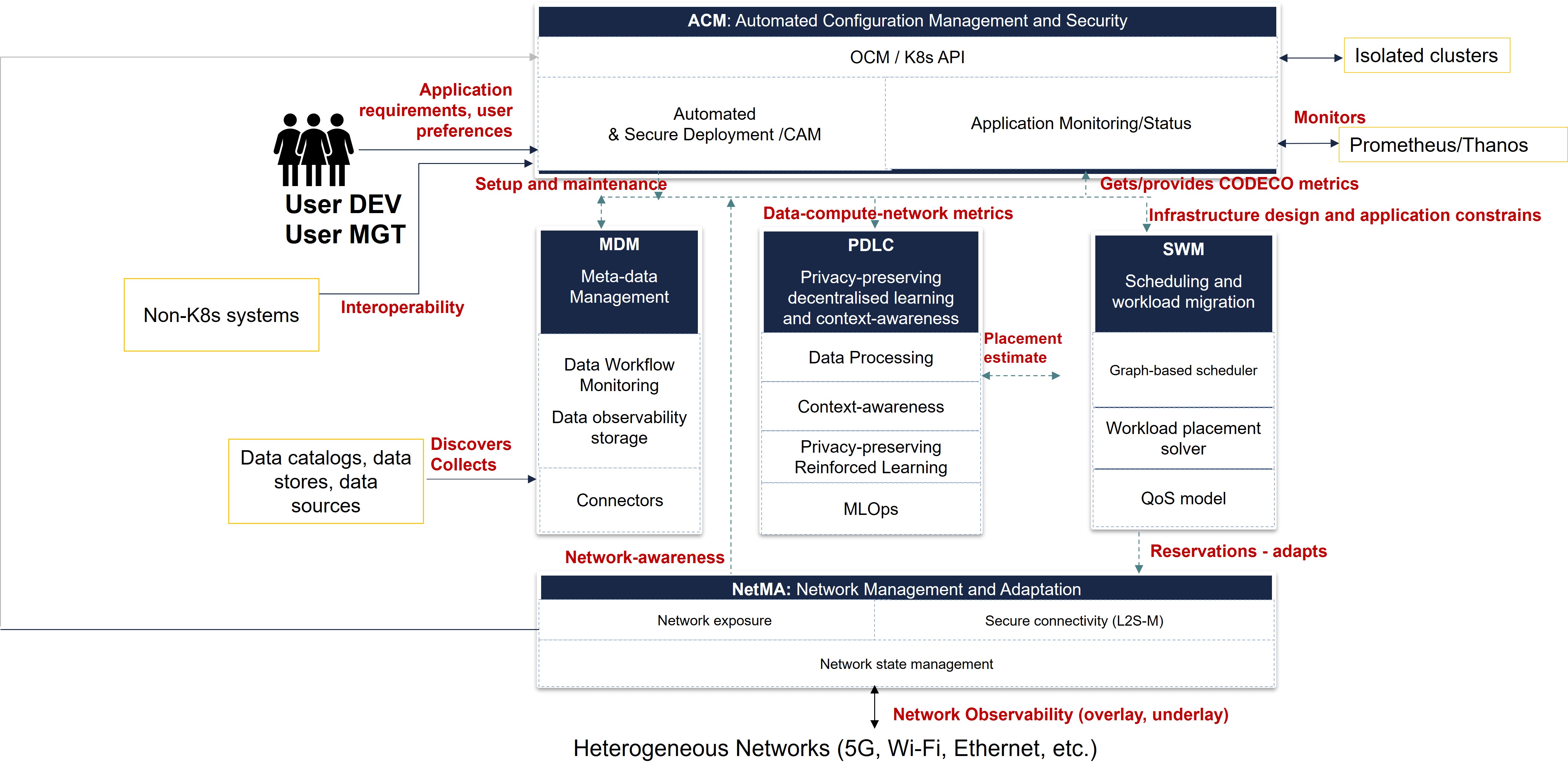}
  \caption{CODECO components and their high-level interactions.}
  \label{fig:codeco-components}
\end{figure*}

\section{CODECO Federated Architecture} 
\label{sec:codeco-fc-arch}
The CODECO federated architecture is designed to support application-centric orchestration across the CEI continuum, where infrastructure
resources span heterogeneous computational, networking, and data environments, often operated by different tenants and administrative domains. Rather than managing clusters as isolated execution units, CODECO treats applications and their microservices as first-class entities that may be distributed, migrated, and adapted across multiple clusters over time.

\subsection{Core Architectural Components}

\subsubsection{Management Functions and Component Mapping} 
Figure~\ref{fig:codeco-components} illustrates the CODECO components and their high-level interactions:

\begin{itemize}
  \item \textbf{ACM-FC (Advanced Configuration and Management FC)} is the   primary user-facing component. It manages application lifecycles, interprets   application intent expressed through a semantic abstraction, the \textit{CODECO Application MOdel (CAM)}, coordinates neighborhood creation,   and interfaces with the underlying federation.
  \item \textbf{SWM-FC (Seamless Workload Migration FC)} is responsible for workload   placement, migration, and re-scheduling across clusters. It enforces scheduling decisions based on application constraints, infrastructure conditions, and recommendations provided by learning components.
  \item \textbf{PDLC-FC (Privacy Decentralised Learning and Context-awareness)} provides AI-assisted, context-aware recommendations. It analyses cross-layer metrics (compute, network, data) and produces cost and stability indicators without directly executing placement decisions.
  \item \textbf{NetMA-FC (Network Management and Adaptation)} supplies network observability, secure inter-cluster connectivity, and network capability exposure required for federated orchestration.
  \item \textbf{MDM-FC (Metadata Management)} maintains a metadata graph capturing data locality, freshness, portability, and compliance constraints, enabling data-aware orchestration decisions.
\end{itemize}

Each component is responsible for a distinct aspect of orchestration, and with specific management roles as provided in Table \ref{tab:codeco-management-functions}.

\begin{table}[htbp]
\centering
\caption{Management functions and component mapping.}
\label{tab:codeco-management-functions}
\small
\renewcommand{\arraystretch}{1.05}
\begin{tabular}{p{2.8cm} p{4.2cm}}
\toprule
\textbf{Management Function} & \textbf{CODECO Components} \\
\midrule
Intent and Policy Management &
ACM-FC (CAM), OCM Hub \\
Scope and Federation Management &
Neighborhood Composer (CNC), ACM-FC Neighborhood Manager \\
Decision Support and Context-Awareness &
PDLC-FC (CA, GNNs, MARL) \\
Scheduling and Migration Enforcement &
SWM-FC (SWM-L1, SWM-L2) \\
Network Configuration and Connectivity &
NetMA-FC (NSM, Nemesys-FC, Secure Connectivity) \\
Data Governance and Data Awareness &
MDM-FC (Metadata Graph, Connectors, API) \\
Observability and Feedback &
ACM-FC, NetMA-FC, MDM-FC \\
\bottomrule
\end{tabular}
\end{table}

\subsection{Management Layers and Decision Timescales}  
CODECO structures management decisions across three temporal layers: (i) a governance layer operating at long timescales to define intent, policy, and federation membership; (ii) a coordination layer operating at medium timescales to scope orchestration through application neighborhoods; and (iii) an execution layer operating at short timescales to enforce placement, migration, and network configuration using fresh local context.

As summarized in Table~\ref{tab:codeco-management-functions}, CODECO components are positioned across these layers according to their functional role and temporal requirements. Hub-spoke components such as ACM-FC and OCM operate predominantly at the governance layer, while PDLC-FC and SWM-FC span coordination and execution layers, enabling adaptive behavior without breaking global policies. 

\subsection{Federation and Control Structure}
\begin{figure*}[t!]
  \centering
  \includegraphics[width=\textwidth]{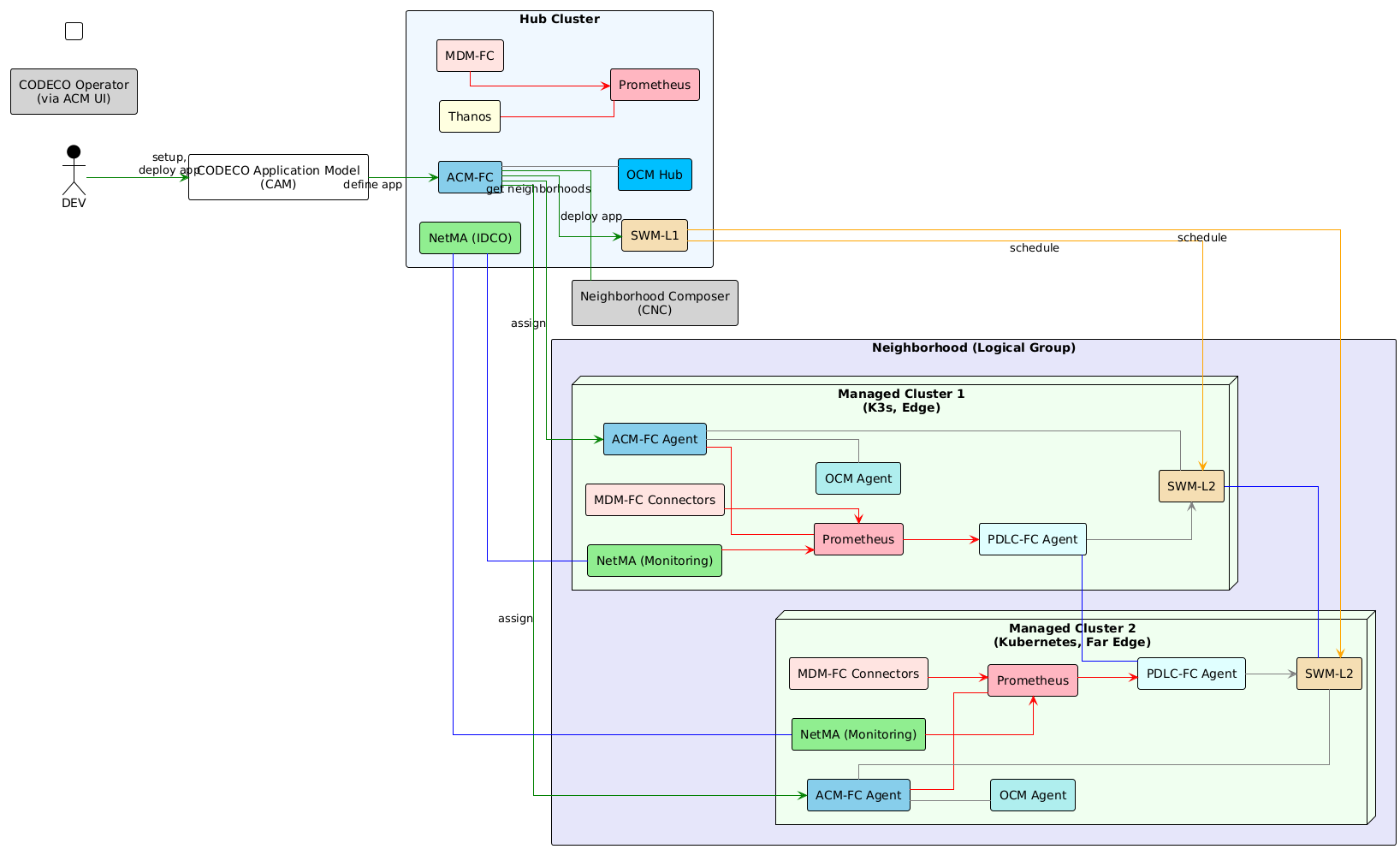}
  \caption{CODECO components, positioning across federated clusters based on the Hub-and-spoke model of OCM. Green arrows represent the user to  CODECO logic; red arrows represent the CODECO observability interfaces. Orange arrows correspond to the Hub to spoke control plan  communication between CODECO components; blue arrows represent the communication between CODECO components across the control plane of MCs.}
  \label{fig:codeco-fc}
\end{figure*}

CODECO relies on OCM as the foundation for federated coordination. A minimal deployment consists of one \textit{Hub} cluster and one or more \textit{Managed Clusters (MCs)}, as shown in Figure~\ref{fig:codeco-fc}. OCM provides cluster registration, identity management, policy distribution, and manifest dispatch, establishing a common governance layer across the federation.

CODECO does not assume continuous, low-latency connectivity between the Hub and all Managed Clusters. To support far Edge environments characterized by mobility and intermittent connectivity, it adopts a hybrid control-plane model. The Hub remains the authoritative source of desired state and policy enforcement, while MCs retain sufficient autonomy to continue microservice placement, re-scheduling, and inter-cluster communication during hub disconnections. This design preserves governance consistency while enabling resilient operation.

\subsection{Application Neighborhoods: Federated Partitioning}

To control complexity and improve scalability, CODECO introduces the concept of \textit{application neighborhoods}. A neighborhood is a bounded, semantic grouping of MCs selected for a specific application based on its requirements, performance objectives, and contextual constraints. Each application is associated with a single neighborhood, while a Managed Cluster may belong to multiple 
Neighborhoods define the coordination scope for scheduling, monitoring data exchange, learning collaboration, and inter-cluster communication. By limiting coordination to application-relevant subsets of clusters, CODECO avoids global state dissemination, reduces overhead, and improves scalability. Neighborhoods are computed at deployment time and are expected to remain stable during normal
operation, although re-evaluation may be triggered by sustained performance degradation, resource depletion, or environmental instability.

To further reduce orchestration overhead, CODECO decouples \textit{space partitioning} from \textit{neighborhood selection}. Similarity-based clustering of MCs can be performed periodically (e.g., offline or periodically) to establish coarse partitions, while application-specific neighborhood selection is restricted to the most relevant subset at deployment time. This separation significantly lowers computational and communication costs while preserving adaptivity to changing conditions.

\subsection{Semantic Application Abstraction and Observability}

Users interact with CODECO exclusively through ACM-FC by submitting applications described using the  CAM, a YAML-based semantic abstraction that specifies microservices, service dependencies, and performance intents such as latency, energy efficiency, resilience, and
compliance. CAM is intentionally agnostic to whether an application is deployed within a single cluster or across multiple clusters, allowing seamless migration from local to federated execution.

The \texttt{spec} section of CAM defines application structure and requirements, while the \texttt{status} section is continuously updated by ACM-FC with runtime observability information. 

\subsection{Federated Component Placement}

CODECO components are distributed across the federation according to their functional roles. Such placement ensures centralized governance while keeping monitoring, learning, and execution close to the resources that each component handles:
\begin{itemize}
  \item \textbf{ACM-FC} runs in the Hub cluster, with lightweight agents deployed
  in each MC.
  \item \textbf{PDLC-FC} instances operate in the control planes of MCs belonging to a neighborhood.
  \item \textbf{SWM-FC} is split into a global coordination layer (SWM-L1) hosted in the Hub and local execution layers (SWM-L2) deployed in each Managed Cluster.
  \item \textbf{NetMA-FC} spans the federation, with monitoring agents close to workloads and inter-domain connectivity orchestrated at the Hub level. 
  \item \textbf{MDM-FC} is typically instantiated per neighborhood to manage application-relevant metadata.
\end{itemize}

\subsection{Component Interaction and AI-Driven Decision Flow}

Once an application is described via CAM and deployed, ACM-FC orchestrates the activation of all CODECO components. Monitoring data from NetMA-FC, MDM-FC, and ACM-FC is collected and exposed through Prometheus. PDLC-FC aggregates these cross-layer observations and produces context-aware cost and stability estimates, which are delivered to SWM-FC as node and cluster recommendations based on specific performance profiles and existing resources.

SWM-FC considers PDLC recommendations along with the application constraints described in the CAM, and with the real-time infrastructure conditions provided by ACM-FC and NetMA-FC. SWM relies on these parameters to ultimately reach an optimal solution to perform placement, migration, and re-scheduling decisions. Network configuration instructions are forwarded to NetMA-FC, which establishes secure overlays and exposes updated network metrics. MDM-FC continuously updates data-related constraints and metadata, ensuring that orchestration decisions respect compliance, locality, and data mobility requirements.

\section{Components' Operational Roles}
\label{Sec: components}

To avoid redundancy with the architectural description in Section~\ref{sec:codeco-fc-arch}, this section focuses exclusively on the CODECO component runtime behavior, inputs, outputs, and component interactions.

\begin{figure*}[h!]
  \centering
  \includegraphics[width=0.9\linewidth]{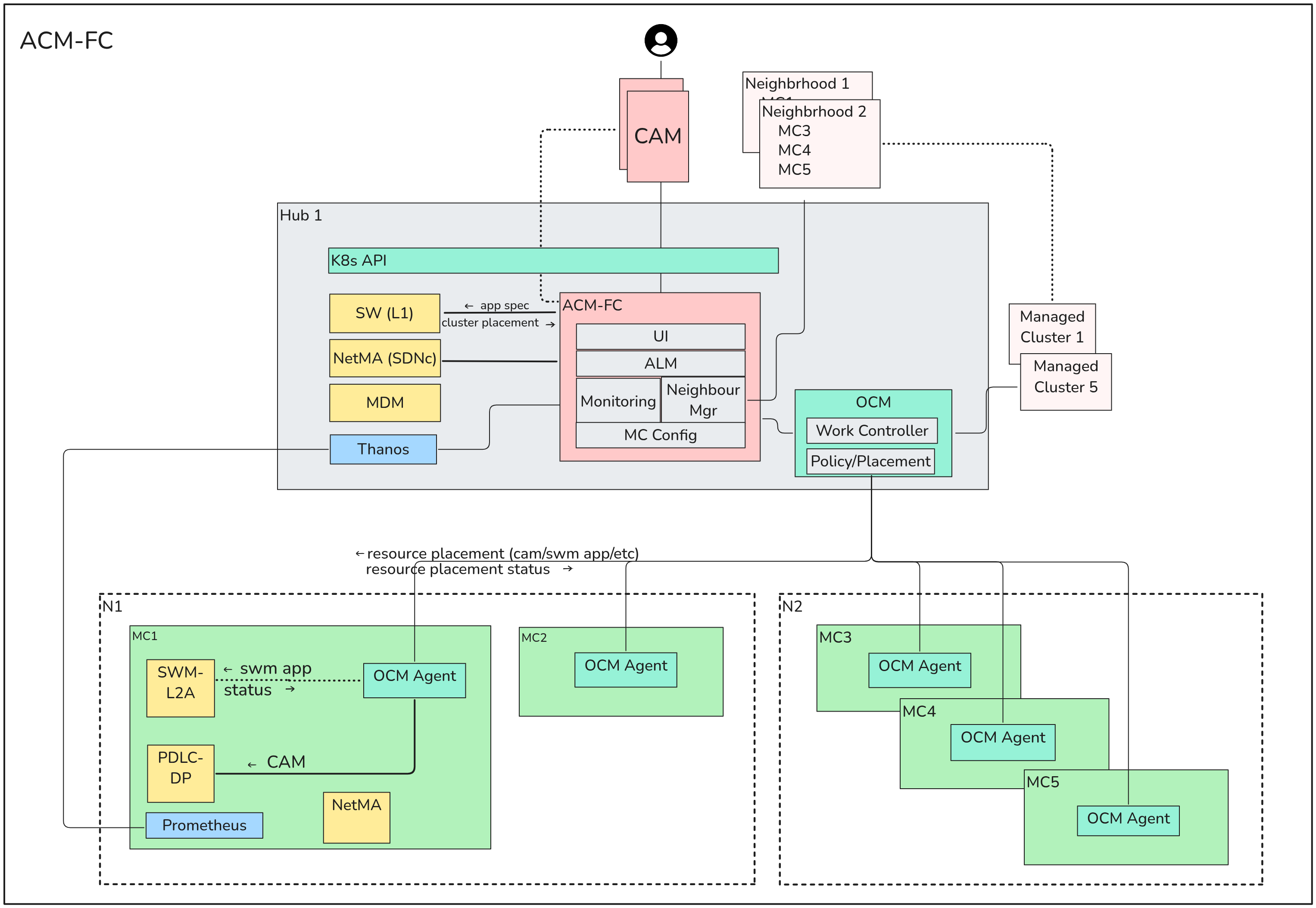}
  \caption{ACM-FC deployment across one Hub and two Managed Clusters.}
  \label{fig:acm-fc}
\end{figure*}

\begin{figure}[ht!]
  \centering
  \includegraphics[width=\columnwidth]{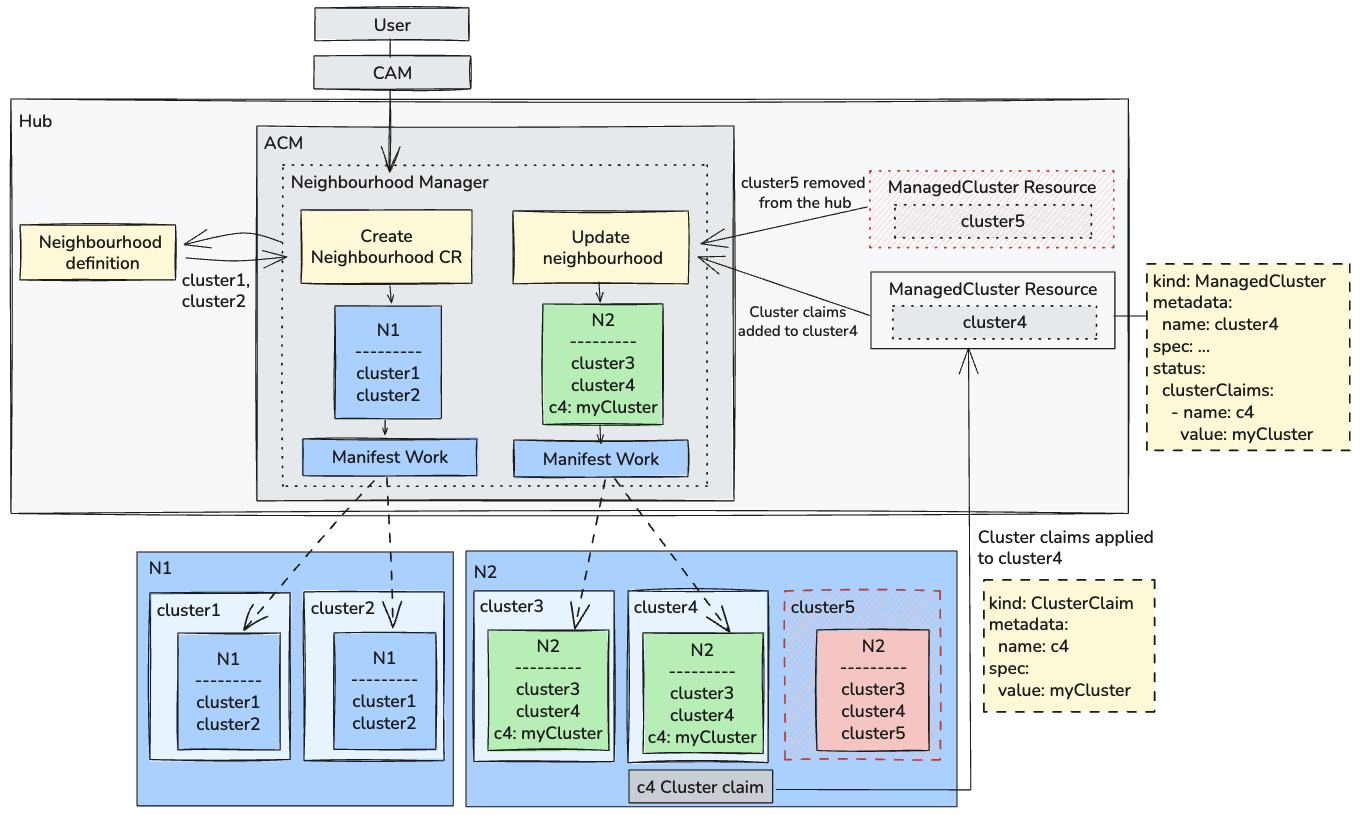}
  \caption{ACM-FC and the neighborhood Manager.}
  \label{fig:acm-nm}
\end{figure}

\subsection{CODECO Neighborhood Composer Plugin: Partitioning the Problem Space}
\label{subsec:cnc}

A key scalability challenge in federated CEI orchestration is limiting the search and coordination space without sacrificing adaptivity. CNC is designed to separate two tasks that otherwise become expensive at scale:
(i) \textbf{space partitioning}, which groups clusters into coarse similarity classes at a slower timescale; and (ii) \textbf{neighborhood selection}, which chooses a bounded set of candidate clusters for a specific application at deployment time. This reduces repeated global optimization and limits monitoring and learning exchanges to application-relevant clusters.

CNC separates coarse-grained cluster partitioning (performed infrequently) from application-specific neighborhood selection at deployment time. Given a CAM and target performance profile, it outputs a bounded neighborhood specification consumed by ACM-FC and disseminated via OCM.

\subsection{ACM-FC}
\label{subsec:acm-fc}

ACM-FC is the governance-facing entry point of CODECO and is primarily deployed at the Hub cluster. In a federated setting, ACM-FC fulfills three main management responsibilities:
\begin{itemize}
    \item \textbf{Platform configuration across the federation}, including onboarding of new
    clusters and enforcement of baseline configuration required by CODECO components.
    \item \textbf{Intent and lifecycle coordination}, by validating the CODECO CAM and translating application intent into internal representations consumed by    scheduling, networking, and learning components.
    \item \textbf{Observability integration}, aggregating monitoring information from the     federation and reflecting application and infrastructure status back into CAM.
\end{itemize}

At deployment time, ACM-FC validates CAM, triggers neighborhood selection (via CNC or
policy-driven mechanisms), and coordinates the activation of CODECO components for the
selected neighborhood. During runtime, ACM-FC maintains application lifecycle state
(updates, rollback, removal) and ensures that monitoring data collected from the federation is exposed to users in a consistent manner.

As illustrated in Figure~\ref{fig:acm-fc}, internally ACM-FC comprises an Application Lifecycle Manager (ALM), Multi-Cluster Configuration (MCC), and a neighborhood Manager, which together reconcile CAMs, deploy CODECO services across clusters, and maintain neighborhood definitions.

\subsubsection{The ACM-FC Neighborhood Manager}
\label{subsubsec:acm-nm}

The \textit{neighborhood manager} operates as an internal sub-component of ACM-FC, as shown in Figure~\ref{fig:acm-nm}. Its primary responsibility is to manage the lifecycle of neighborhood definitions associated with applications.

Once a neighborhood is produced (by CNC or by policy), the neighborhood Manager creates a corresponding Kubernetes resource capturing membership and constraints and disseminates it to the relevant MCs through OCM \textit{ManifestWork}. This enables other CODECO components
(e.g., PDLC-FC and SWM-FC) to discover neighborhood membership locally without requiring continuous interaction with the Hub cluster.

The neighborhood Manager maintains neighborhood consistency over time through reconciliation.
Cluster lifecycle events (join or leave) and relevant metadata changes (e.g., labels or claims)
can trigger updates to the neighborhood resource and its redistribution. By isolating
neighborhood handling within ACM-FC, CODECO avoids embedding federation scoping logic
directly into scheduling or learning components\footnote{In early experimentation phases, neighborhood membership may be configured statically to validate workflows. Full CNC-driven neighborhood selection is integrated progressively as part of the CODECO open-source toolkit.}.
\label{PDLC-FC}
\begin{figure*}[h!]
  \centering
  \includegraphics[width=0.7\textwidth]{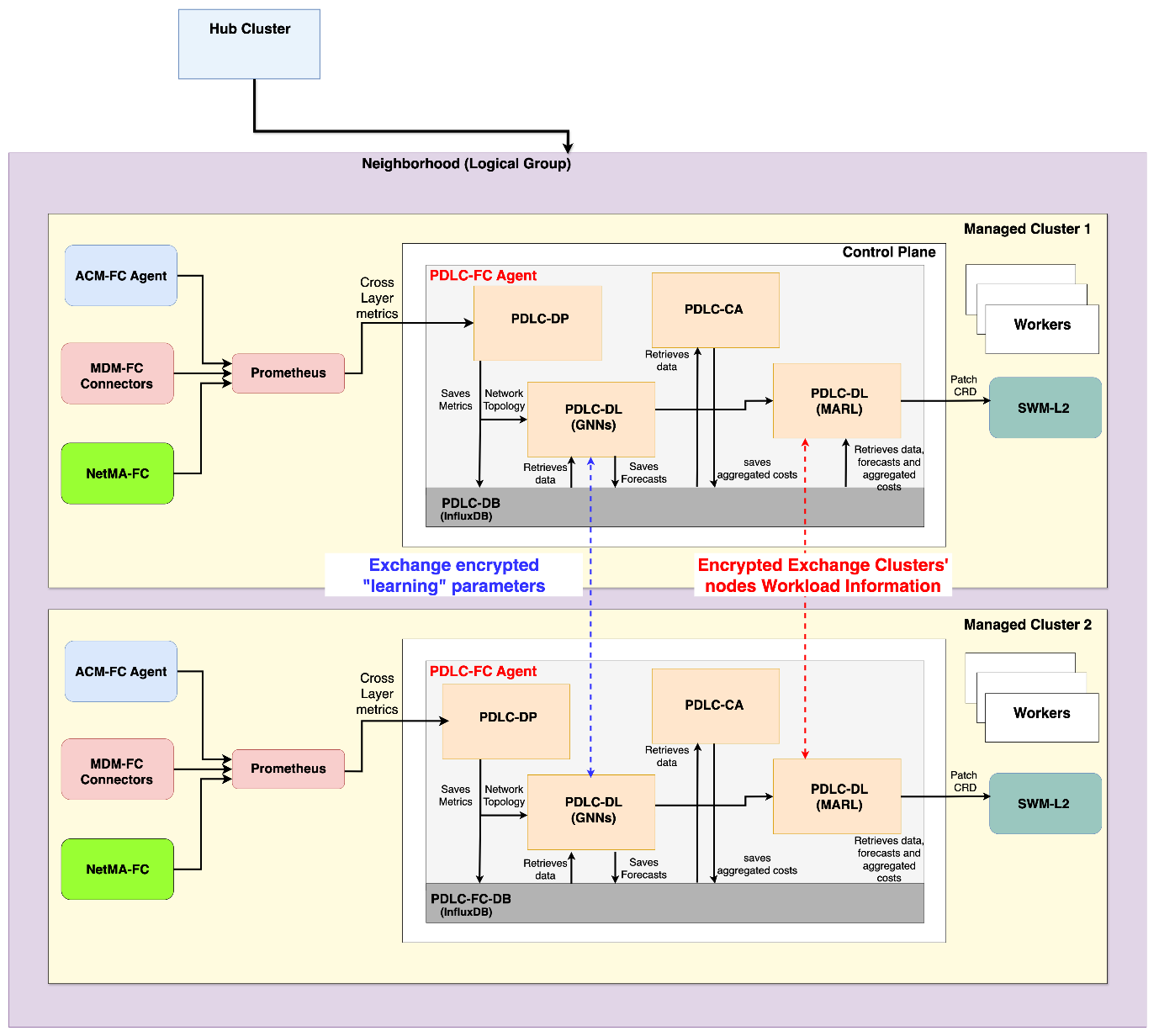}
  \caption{Functional, high-level representation of the PDLC sub-components across a federated environment and of their interactions.}
  \label{fig:pdlc-fc}
\end{figure*}

\subsection{PDLC-FC}
PDLC-FC provides AI-assisted decision support for federated orchestration by analyzing cross-layer observability data and producing placement and migration recommendations. It operates in a decentralized and privacy-preserving manner, keeping raw telemetry local to each Managed Cluster and exchanging only protected learning artefacts within application neighborhoods. PDLC-FC does not enforce decisions; instead, it supplies cost and stability indicators that are consumed by SWM-FC.

A functional overview of PDLC-FC is given in Figure~\ref{fig:pdlc-fc}. CODECO enforces a clear separation between \emph{recommendation} and \emph{enforcement}: PDLC-FC produces advisory outputs, while scheduling and re-scheduling actions for micro-services (workloads) remain the responsibility of SWM-FC. Decentralized learning is therefore not used as the main scheduling mechanism. This decoupling improves interoperability with Kubernetes-aware schedulers and avoids retraining scheduling logic when new objectives or constraints are introduced.

In federated operation, PDLC-FC instances run in MCs and interact only with peer instances belonging to the same application neighborhood. This bounded collaboration aligns with CODECO’s privacy, security, and data-sovereignty constraints
and limits communication overhead.

PDLC-FC operates a continuous recommendation loop that aggregates local compute, network, and data metrics into cost and stability indicators, optionally enhanced with short-term forecasts, which are exposed via Kubernetes resources.

\subsubsection{PDLC-DP: Data Processing}
PDLC-DP is responsible for retrieving cross-layer metrics from Prometheus, normalizing them, and maintaining a local telemetry store for PDLC sub-components. Each PDLC-FC instance operates exclusively on locally collected data, avoiding shared storage dependencies and preserving data sovereignty. PDLC-DP also retrieves neighborhood membership information to scope learning collaboration and recommendation dissemination.

\subsubsection{PDLC-CA: Context Scoring}
\begin{figure}[ht!]
\centering 
\includegraphics[width=\columnwidth]{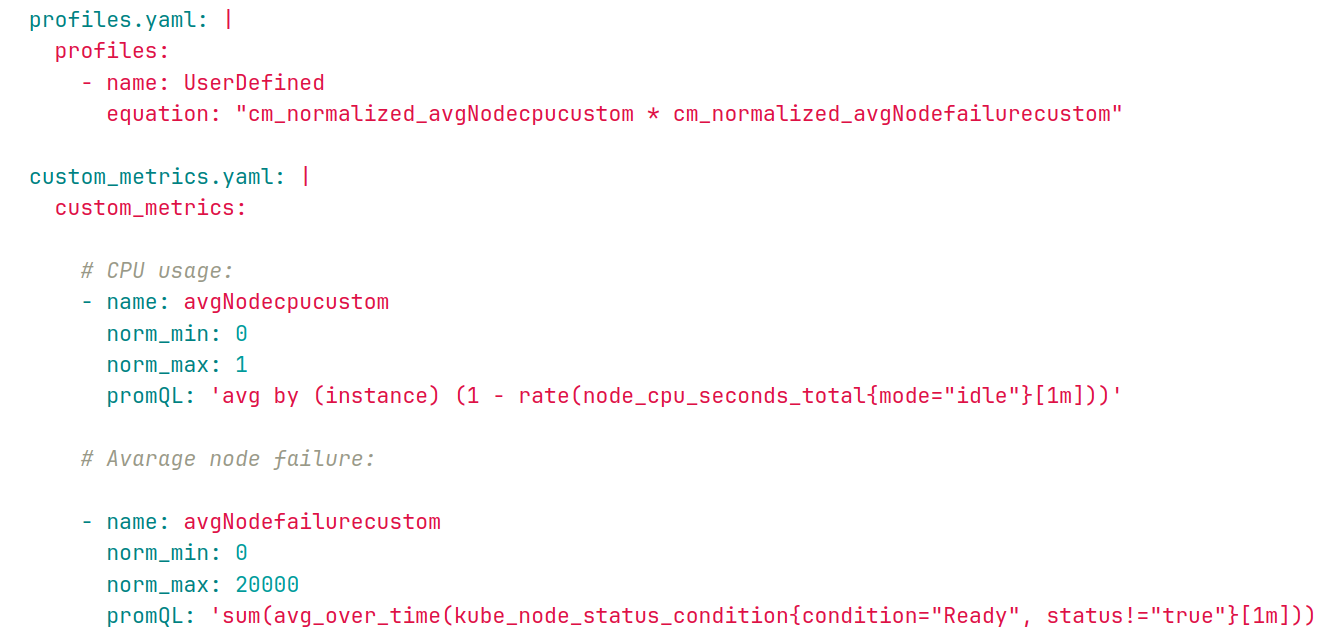} \caption{Example for user-defined target performance profiles and weights.} 
\label{fig:pdlc-fc-custom} 
\end{figure}

\begin{figure}[ht!] 
\centering 
\includegraphics[width=\columnwidth]{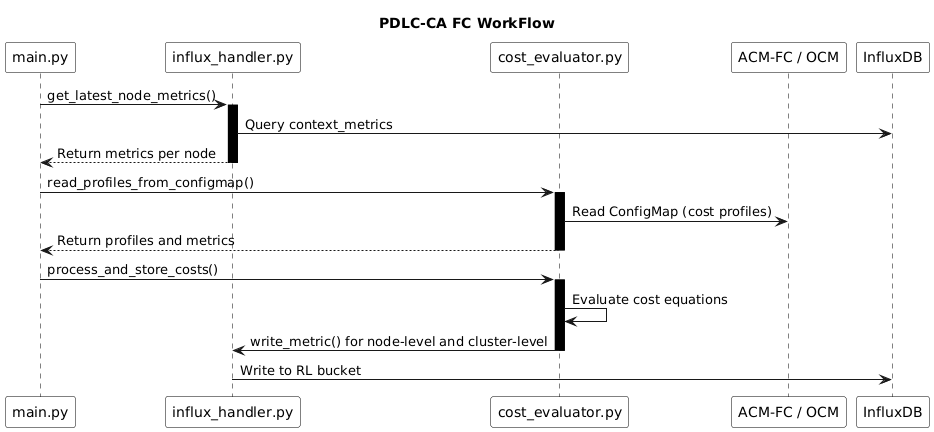}\caption{PDLC-CA sequence diagram.} \label{fig:pdlc-ca-sequence} 
\end{figure}

PDLC-CA supports predefined and user-defined performance profiles by aggregating normalized metrics into compact node and cluster scores based on the user defined target profiles, e.g., Greenness, Resilience (rf. to section \ref{sub:context-awareness}). In addition, developers may define custom profiles by mixing normalized CODECO metrics with newly registered custom metrics. Figure~\ref{fig:pdlc-fc-custom} illustrates how such custom definitions are structured and referenced within PDLC-CA-FC. These indicators are exposed
to downstream components, including PDLC-DL and SWM-FC, to support context-aware
placement and migration decisions without modifying core scheduling logic, as shown in Figure igure~\ref{fig:pdlc-ca-sequence}.

\subsubsection{PDLC-GNN: Infrastructure Forecasting}

The PDLC-DL-GNNs sub-component forecasts short-horizon infrastructure utilization to support proactive placement and migration decisions. Each PDLC instance embeds a local GNN module and operates independently within its cluster control plane. The model follows a spatio-temporal approach: it captures the evolution of node metrics over time while accounting for structural relationships between nodes, enabling topology-aware predictions.

\begin{figure*}[t!]
\centering
\includegraphics[width=0.8\textwidth]{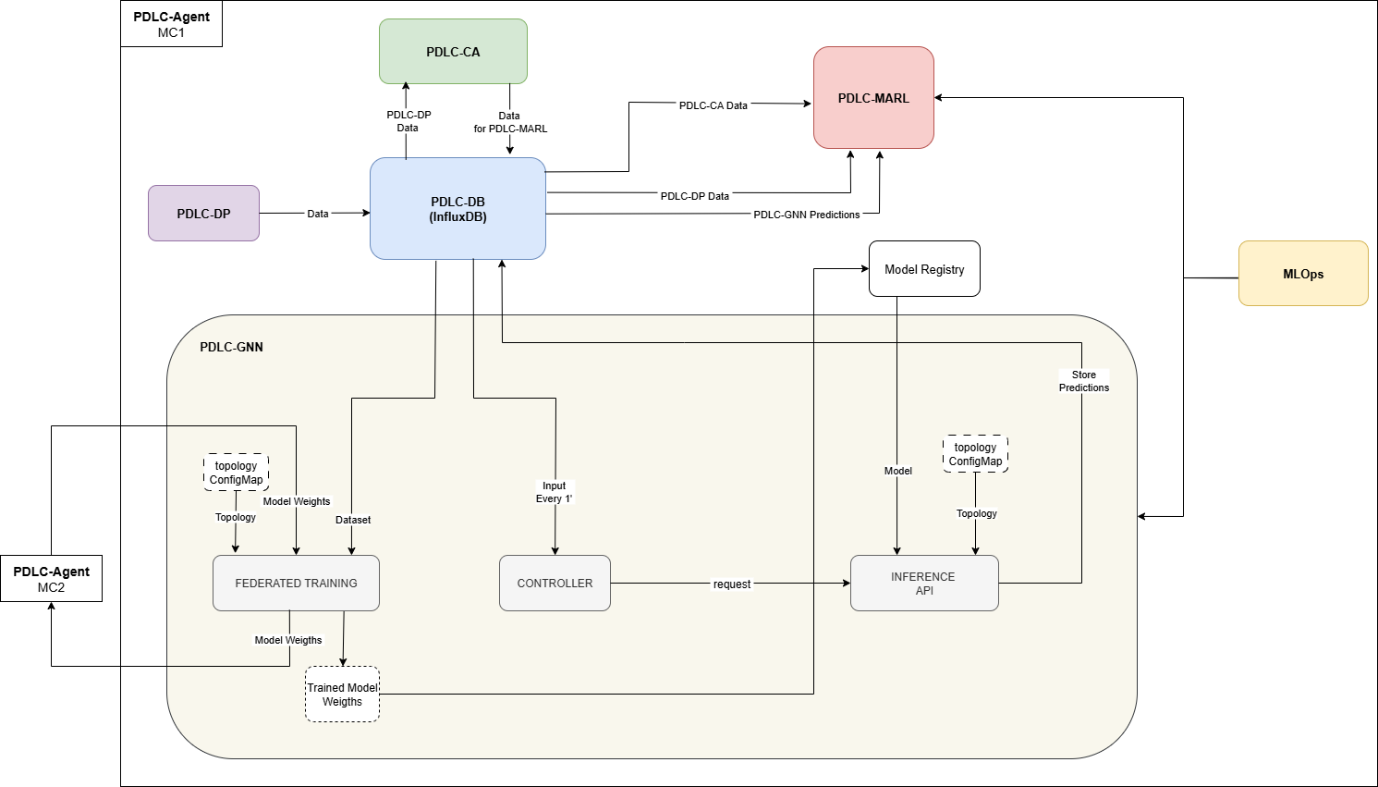}
\caption{PDLC-GNN high-level representation.}
\label{fig:pdlc-gnn}
\end{figure*}

PDLC-DL-GNNs is organized around three main elements as represented in Figure~\ref{fig:pdlc-gnn}):
\begin{itemize}
    \item \textbf{Decentralized federated training:} each cluster trains its GNN locally using metrics stored by PDLC-DP. At defined intervals, clusters exchange encrypted model weights with neighborhood peers and aggregate them securely, improving model quality without sharing raw telemetry.
    \item \textbf{Controller:} orchestrates inference cycles by retrieving recent metrics from the local database, assembling model inputs, and tracking node membership to reflect the active topology.
    \item \textbf{Inference API:} loads the appropriate model variant and produces short-term orecasts (e.g., 5/15/30 minutes), which are stored locally for consumption by PDLC-DL-MARL.
\end{itemize}

Models are maintained in a local registry. If the topology changes (nodes join/leave), PDLC selects or updates an appropriate model, and may trigger re-training when mismatch or degradation is detected.

\subsubsection{PDLC-DL-MARL}
\begin{figure*}[t!]
\centering
\includegraphics[width=0.6\textwidth]{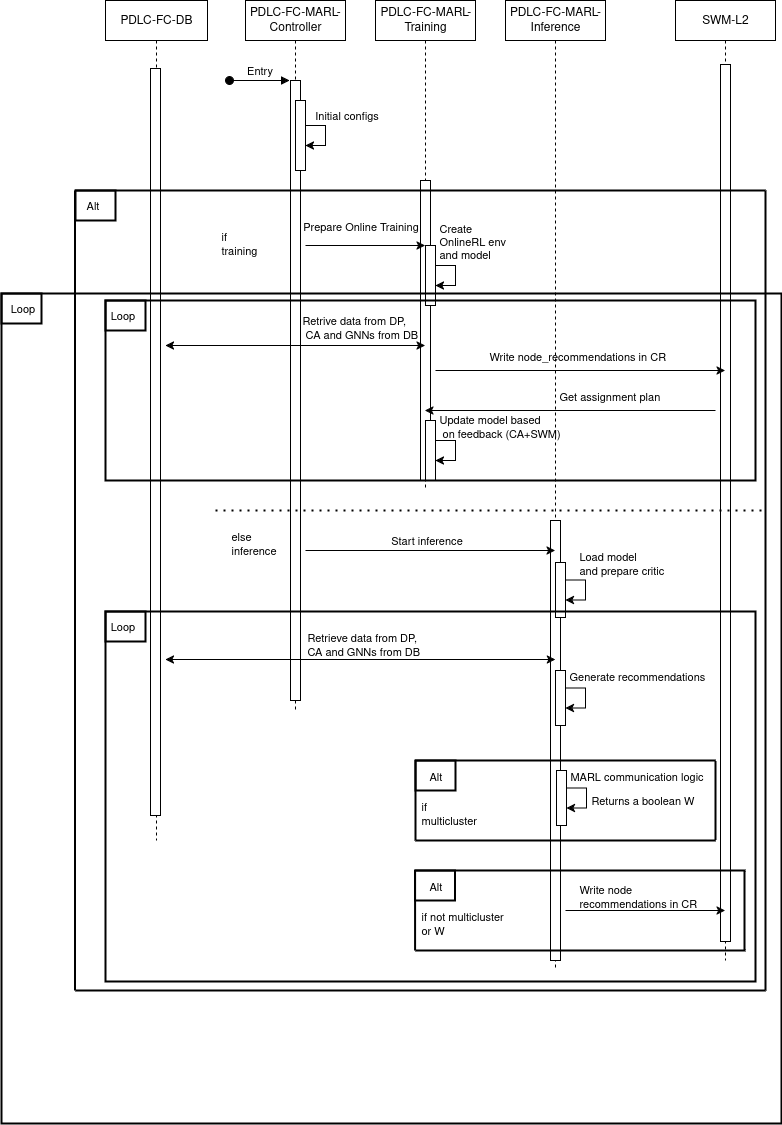}
\caption{PDLC-DL-MARL communication sequence.}
\label{fig:pdlc-marl-architecture}
\end{figure*}

PDLC-DL-MARL enables neighborhood-scoped inter-cluster recommendation through a multi-agent reinforcement learning (MARL) design. Figure~\ref{fig:pdlc-marl-architecture} illustrates the interaction sequence. Agents run in the control plane of MCs and operate without a central coordinator. Rather than exchanging raw state or full model parameters, agents share high-level decision values, supporting federated collaboration consistent with CODECO security requirements.

PDLC-DL-MARL comprises (i) a \textit{data ingestion layer} reading observations from the local InfluxDB instance, (ii) an \textit{RL controller} managing training/inference modes and communication/security checks, and (iii) an \textit{inter-agent coordination layer} based on an auction-style mechanism where clusters compete to host workloads by exchanging decision values.

The training controller follows the RL loop and includes: (i) \textbf{topology resilience} through padding to generalize across clusters of varying sizes (up to 20 nodes), and (ii) \textbf{action masking} to filter infeasible actions at runtime. Observations are retrieved from InfluxDB and recommendations are written to the appropriate custom resources.

PDLC-DL-MARL enables decentralized coordination among Managed Clusters within an
application neighborhood. Each cluster runs a local agent that evaluates its
suitability for hosting workloads based on observed context and predicted
conditions. Agents exchange abstract decision values rather than raw state or
model parameters, enabling cooperative recommendation while preserving privacy
and limiting communication overhead.

\subsubsection{PDLC Functional Coverage of Security, Resilience, and Scalability}

PDLC-FC inherently supports security, resilience, and scalability through its
decentralized design. Telemetry remains local to each cluster, inter-cluster
exchanges are protected, and collaboration is limited to application
neighborhoods. As a result, failures or instability in individual clusters do not
propagate, and learning-related communication remains bounded as the federation
grows.

\subsection{NetMA-FC}
\label{netma-fc}
NetMA-FC provides network observability, secure inter-cluster connectivity, and
capability exposure required for federated CEI orchestration. It enables CODECO
to incorporate network latency, bandwidth, loss, and energy indicators as
first-class inputs to placement and adaptation decisions.

A functional overview of NetMA-FC, its sub-components, and its interfaces to the remaining CODECO stack is shown in Figure~\ref{fig:netma-fc}. NetMA-FC comprises three main sub-components:

\begin{figure*}[t!]
\centering
\includegraphics[width=0.8\textwidth]{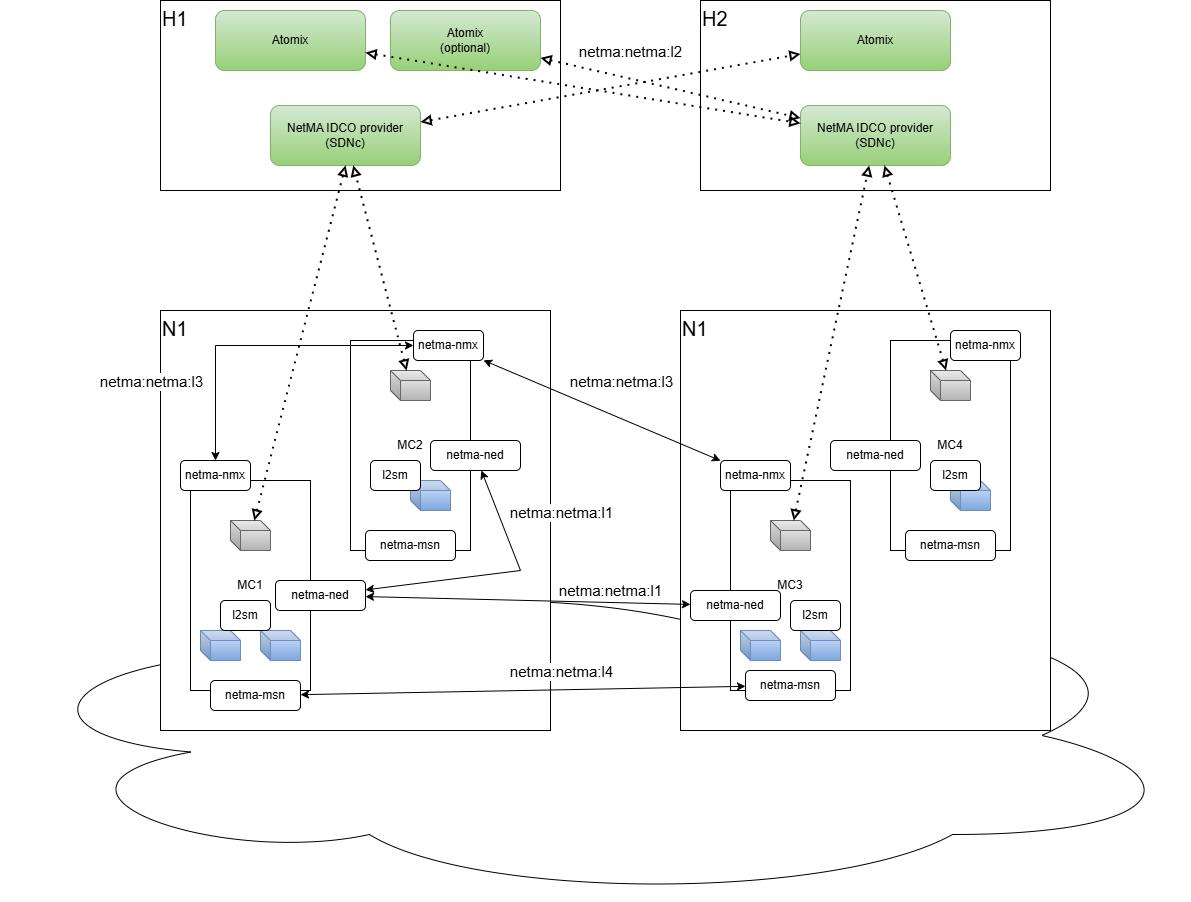}
\caption{NetMA sub-components placement across hub clusters H1, H2, and MC clusters N1, N2.}
\label{fig:netma-fc}
\end{figure*}

\begin{itemize}
  \item \textbf{Network State Monitoring (NSM)} provides underlay and overlay observability,   capturing link- and path-level conditions relevant to application QoS and mobility.
  \item \textbf{Network Exposure System (Nemesys-FC)} acts as the NetMA-FC northbound interface,   translating monitored information into consumable cost maps and metrics, and exposing them via   Prometheus/Kubernetes resources to ACM-FC, PDLC-FC, and SWM-FC.
  \item \textbf{Secure Connectivity (L2S-M)} establishes secure and isolated inter-cluster   communication for CODECO workloads. In the federated phase, this includes a distributed   SDN control plane for inter-domain connectivity and support for both pre-established and on-demand overlay topologies.
\end{itemize}

\subsubsection{NSM: Network State Monitoring}
NSM provides underlay and overlay network observability at the Kubernetes
worker-node level, where application execution context and packet visibility
coexist. It collects metrics such as latency, bandwidth, packet loss, link
failures, and link/flow energy consumption, enabling application-aware and energy-aware
network assessment in federated environments.
\paragraph{Underlay and Overlay Observability}
NSM performs underlay and overlay monitoring at Kubernetes worker nodes using eBPF-based instrumentation. Metrics include latency, bandwidth, loss, link failures, and energy consumption, enabling application-aware and energy-aware network assessment. Metrics are exported to Prometheus and consumed by PDLC-FC and SWM-FC.

The \textit{MON} module (netma-nsm-mon) is a sub-component of NetMA responsible for source-node passive network probing and aggregation of underlay metrics. MON builds on the open-source Kubernetes \emph{k8s-netperf}\footnote{\url{https://github.com/leannetworking/k8s-netperf.}} plugin, extended to support additional metrics including energy-related ones. MON uses eBPF instrumentation to monitor the parameters shown in Table~\ref{tab:netma-nsm-mon-metrics} and adopts, for energy-aware metrics (link and flow level) the linear power model proposed by Feeney~et~al.~\cite{feeney2001}. This model estimates packet transmission power as a linear function of packet size, combining a static baseline power with a size-dependent term. The total link energy over a monitoring period is obtained by accumulating the energy consumed by individual packet transmissions, accounting for their respective transmission durations. Other formulations can be used.

\begin{table*}[t]
\centering
\caption{Network metrics exported by \texttt{MON}.}
\label{tab:netma-nsm-mon-metrics}
\small
\setlength{\tabcolsep}{6pt}
\renewcommand{\arraystretch}{1.15}
\begin{tabular}{p{3cm} p{6.8cm} p{2.2cm} p{2cm}}
\hline
\textbf{ID (Attribute)} & \textbf{Description} & \textbf{Values} & \textbf{Units} \\
\hline
\texttt{nodeName} & Identifier of the node in Kubernetes & string & n.a. \\
\texttt{name} (link) & Identifier of a node link & string & n.a. \\
\hline
\texttt{uLinkFailure} & Number of failures of a link & integer (string) & n.a. \\
\texttt{uPacketLoss} & Packet losses in a connection & float (string) & n.a. \\
\texttt{uNodeNetFailure} & Sum of link and endpoint failures & integer (string) & n.a. \\
\texttt{uNodeBandwidth} & Total egress bandwidth across all node links & integer (string) & Bytes (KB, MB) \\
\texttt{uNodeDegree} & Number of active links connected to the node & integer (string) & n.a. \\
\texttt{uLatencyNanos} & Egress link latency (EMA) & float (string) & ns \\
\texttt{uFlowEnergy} & Flow energy expenditure (EMA) & float (string) & J \\
\texttt{uLinkEnergy} & Link energy expenditure (EMA) & float (string) & J \\
\hline
\end{tabular}
\end{table*}

At an overlay level, NSM also supports overlay-specific measurements that quantify tunnel/path behavior (e.g., RTT and loss over encapsulated paths), which is necessary to evaluate secure inter-cluster connectivity and to support path-aware orchestration decisions.

\paragraph{Nemesys-FC: Federated exposure and unified network view}
Nemesys-FC is the \textit{North-Bound Interface (NBI)} of NetMA-FC. It gathers topology and performance data produced by NSM (underlay and overlay), normalizes and formats it, and exposes it to other CODECO components (notably ACM-FC and PDLC-FC) through Prometheus metrics and Kubernetes resources.

\begin{figure}[t!]
  \centering
  \includegraphics[width=\columnwidth]{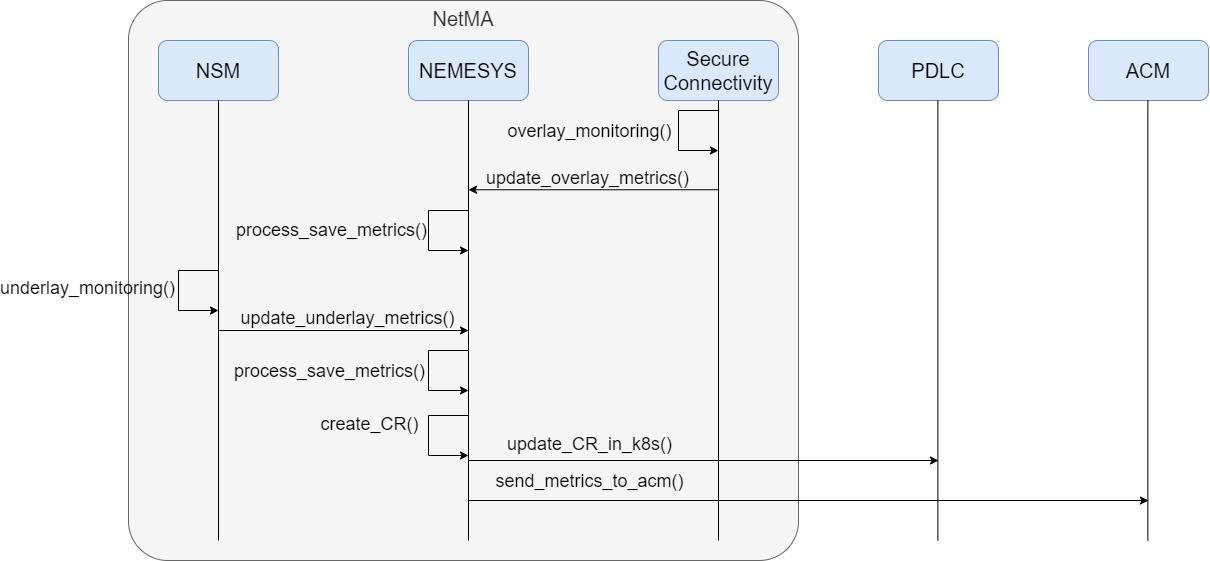}
  \caption{Nemesys interaction with NetMA sub-components, and CODECO components.}
  \label{fig:nemesys-fc}
\end{figure}

Nemesys-FC aggregates underlay and overlay measurements into a unified,
neighborhood-scoped network view and exposes it through Kubernetes resources and
Prometheus metrics. It supports ALTO-based cost maps that represent latency,
bandwidth, loss, and energy, enabling multi-objective scheduling and adaptation
by SWM-FC and PDLC-FC. The interaction of Nemesys to other NetMA sub-components is provided in Figure \ref{fig:nemesys-fc}.

Nemesys-FC follows a modular architecture with (i) underlay and overlay ingestion modules (NSM), (ii) an ALTO core that maintains a local multi-metric graph, and (iii) a federation interface that exchanges neighborhood-scoped summaries. The ALTO core retains the local view and augments it with peer information received through the federation interface, then formats the resulting cost maps into Kubernetes resources and Prometheus metrics. This design preserves local autonomy while enabling cooperative network awareness.

\subsubsection{Secure Connectivity with L2S-M}
\begin{figure*}[t!]
  \centering
  \includegraphics[width=0.8\textwidth]{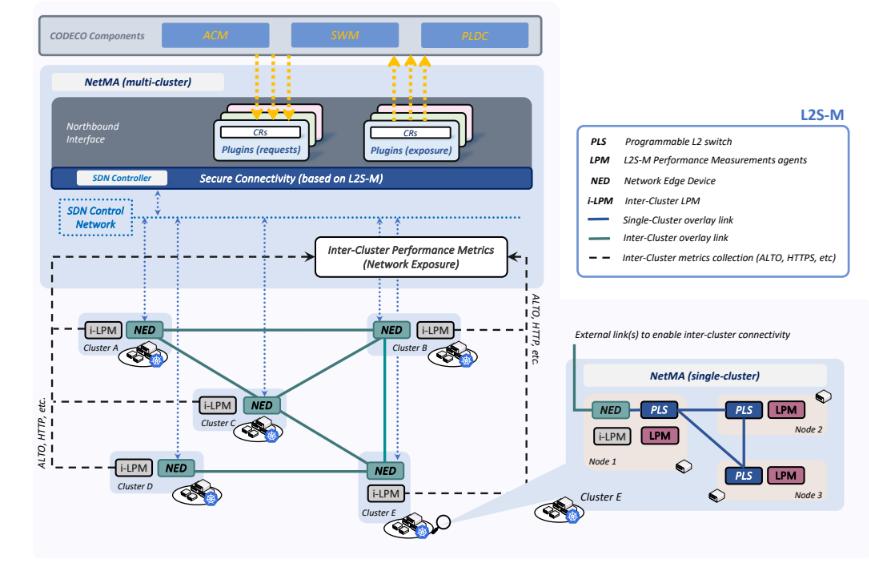}
  \caption{NetMA secure connectivity design.}
  \label{fig:l2sm-fc}
\end{figure*}

The Secure Connectivity sub-component represented in Figure \ref{fig:l2sm-fc} establishes isolated and secure inter-cluster communication for CODECO workloads. It supports both pre-created and on-demand overlay topologies and integrates with federated scheduling decisions to ensure that network paths align with placement outcomes. Continuous monitoring of inter-cluster links enables runtime adaptation under changing
conditions.

\paragraph{Distributed SDN control plane}
To preserve flexibility while aligning with federated operation, Secure Connectivity introduces a distributed SDN controller built on ONOS, deployed as a separate instance in each hub cluster.
Each instance acts as an \textit{Inter-Domain Connectivity Orchestrator (IDCO)} provider and coordinates with peers using the RAFT consensus  algorithm \cite{raft}.

\paragraph{Virtual links, intents, and flow programming}
Secure Connectivity represents secure inter-cluster connections as virtual links (\textit{Vlinks}). Vlinks are compiled into controller intents and then into concrete forwarding rules installed on Network Edge Devices (NEDs) (e.g., via OpenFlow). Path selection can be influenced by SWM-FC decisions, so that secure connectivity follows the orchestration outcome rather than enforcing fixed paths.

\paragraph{Mapping to neighborhoods}
In the federated model, NEDs are managed by hub-cluster SDN controllers in a many-to-one manner (one controller manages multiple NEDs), while neighborhood-to-controller mapping is one-to-one (one controller instance manages all NEDs within a neighborhood). This preserves bounded coordination and avoids global controller coupling across unrelated applications.

\paragraph{Failover and continuity}
To improve network reliability, failover mechanisms allow NEDs to re-associate with another active controller instance if a hub controller becomes unavailable, preserving secure overlay operation under partial failures.

\paragraph{Inter-cluster overlay setup}
Inter-cluster connectivity relies on an overlay network that interconnects clusters and attaches to intra-cluster secure overlays. Two alternatives are supported:

\begin{itemize}
  \item \textbf{Pre-created overlay topology:} inter-cluster tunnels (e.g., VXLAN) are established   at installation time according to a planned topology; link monitoring starts immediately.
  \item \textbf{On-demand overlay topology:} inter-cluster tunnels are created only when a workload requires secure cross-cluster communication. In this approach, workload deployment triggers updates to a Kubernetes resource (e.g., ConfigMap or a dedicated CR), and NED-side watchers configure or tear down tunnels accordingly\footnote{In the current prototype, on-demand tunnel control is under design; the final implementation may use a dedicated CRD instead of a ConfigMap
  to provide stronger lifecycle guarantees.}.
\end{itemize}
Pre-created topologies reduce time-to-connect at runtime but require more elaborate planning and installation; on-demand topologies increase flexibility but introduce additional control-plane logic and API design requirements.

\paragraph{Inter-cluster link performance monitoring (i-LPM).}
The \textit{Link Performance Monitoring (LPM)} functionality is extended to inter-cluster scenarios (i-LPM) to measure overlay tunnel performance (e.g., throughput, RTT, loss) and provide continuous visibility into the reliability of inter-cluster links. i-LPM remains tool-agnostic (with initial support via \textit{iperf} and \textit{ping}) and exports metrics to Prometheus.
To support dynamic (on-demand) overlays, i-LPM endpoints are configured through a Kubernetes resource (e.g., ConfigMap), enabling reconfiguration without redeployment. i-LPM pods are co-located with NEDs to ensure accurate and low-latency measurement.

\paragraph{Inter-cluster monitoring evolution}
NSM was initially conceived for intra-cluster measurements between microservices using a
lightweight socket-based DaemonSet. This provides real-time monitoring of throughput, latency, and packet loss with minimal overhead and integrates naturally with Prometheus for consumption by PDLC-FC and SWM-FC. In the federated phase, NetMA-FC generalizes this monitoring model to inter-cluster overlays by combining NSM telemetry, Nemesys-FC exposure, and i-LPM tunnel measurements within neighborhood-scoped coordination domains.

\subsection{SWM-FC: Scheduling and Workload Migration}
\label{sec:swm}

SWM-FC is responsible for the initial placement, continuous re-scheduling, and, when required, migration of application workloads across federated clusters, while enforcing application QoS constraints. It supports CEI-relevant optimization objectives such as low latency, reduced energy consumption, data sensitivity, and quality of experience (QoE) by combining context-aware observability inputs with node- and cluster-level recommendations produced by PDLC-FC.

SWM-FC exposes a set of Kubernetes \textit{Custom Resource Definitions (CRDs)}, including \texttt{ApplicationGroup}, \texttt{Application}, \texttt{Workload}, \texttt{Channel}, and \texttt{AssignmentPlan}. It does not collect infrastructure telemetry directly; instead, it consumes (i) topology and connectivity updates from NetMA-FC and (ii) placement recommendations from PDLC-FC. Changes in network conditions (e.g., reduced inter-cluster bandwidth) may invalidate existing placements and trigger re-scheduling or migration. Coordination between the Hub and Managed Cluster (MC) instances relies on OCM/ACM mechanisms, while all scheduling logic and optimization remain internal to SWM-FC.

In federated operation, workloads and their communication channels may span multiple MCs within an application neighborhood. SWM-FC addresses three core challenges: (i) synchronizing scheduling-relevant context across clusters, (ii) exposing only abstracted cluster views to preserve scalability and privacy, and (iii) maintaining up-to-date inter-cluster connectivity information to satisfy latency and loss constraints.

Functionally, SWM-FC assigns application workloads to execution nodes and maps \textit{Channels} to network paths that meet compute and communication requirements (e.g., CPU, memory, bandwidth, latency, and service class). Unlike standard Kubernetes schedulers, SWM-FC supports cross-cluster placement and migration while remaining Kubernetes-consistent: it migrates pods and their associated network connections but does not manage application state, which continues to be handled by native application and Kubernetes mechanisms.

\subsubsection{Deployment and Roles of SWM-FC instances}
\begin{figure}[t]
  \centering
  \includegraphics[width=0.8\columnwidth]{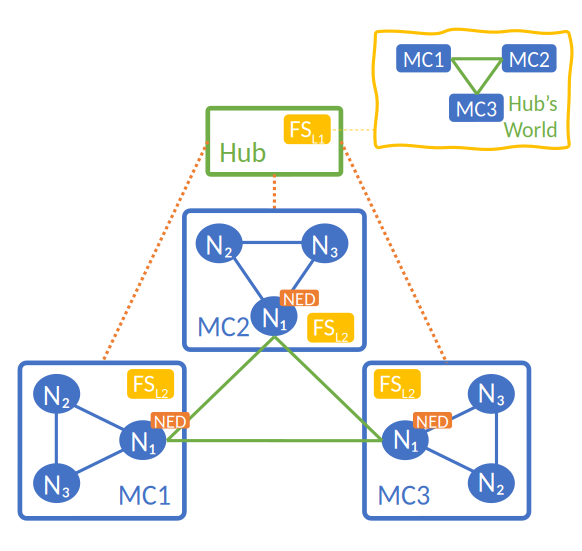}
  \caption{High-level SWM-FC overview: hub-coordinated federated scheduling across
  neighborhood clusters, with per-cluster solvers (FSx) and local execution. Nx represents a worker node.}
  \label{fig:swm-fc-overview}
\end{figure}

SWM-FC runs one instance per cluster: one in the Hub cluster and one in each MC as represented in Figure \ref{fig:swm-fc-overview}.
All instances run the same software, but the Hub instance coordinates multi-cluster placements. MC instances (i) solve placement sub-problems for
their local cluster and (ii) execute the local portion of an agreed placement.
Multi-cluster requests received in an MC are forwarded to the Hub coordinator.

\subsubsection{Two-layer Federated Scheduling model}
\label{subsubsec:swm-fc-model}

SWM-FC assumes that the target cluster set for an application group is the provided CODECO \emph{neighborhood}. The neighborhood is provided by ACM-FC to all CODECO components. neighborhoods are treated as fixed
once scheduling for an application group starts, to avoid ambiguity and
oscillation during placement.

Conceptually, federated scheduling can be decomposed into a layer 1 (L1) problem that distributes workloads across clusters using aggregated cluster information (clusters treated as opaque), and
layer 2 (L2) problems that place workloads on nodes \emph{within} clusters.
While considering two solvability layers, L1 is not solved as a separate optimization stage. Instead, SWM-FC directly uses the neighborhood and then ranks clusters for a sequential search, then solves a series of L2 problems, one per visited
cluster. This avoids “guessing” at L1 from coarse aggregates and reduces unnecessary inter-cluster channels that can arise when clusters are treated as worker nodes. The ordering prefers (a) clusters for which node recommendations are available, and (b) faster inter-cluster links earlier in the sequence.

SWM-FC uses inter-cluster topology resources (e.g., \texttt{NetworkTopology} provided by NetMA) that describe links between clusters. Inter-cluster networks are restricted to single-hop connectivity: two clusters are considered connected only if there is a direct link between them for the relevant service class. Multi-hop routing (e.g., B$\rightarrow$A$\rightarrow$C) is intentionally not used for inter-cluster channels to keep the model tractable and aligned with the available connectivity
substrate.

\subsubsection{Scheduling Workflow}
\label{subsubsec:swm-fc-workflow}

SWM-FC assumes that (i) CODECO applications may be deployed across clusters, (ii) each cluster has a gateway/NED role for inter-cluster traffic, and (iii) interworking with NetMA (including L2S-M) provides the service classes required for assured channels. The detailed mechanisms for Hub-MC communication and the final integration points with ACM/OCM are implementation-dependent and are
treated as part of the CODECO control-plane integration work.

\begin{figure}[ht!]
  \centering \includegraphics[width=0.8\columnwidth]{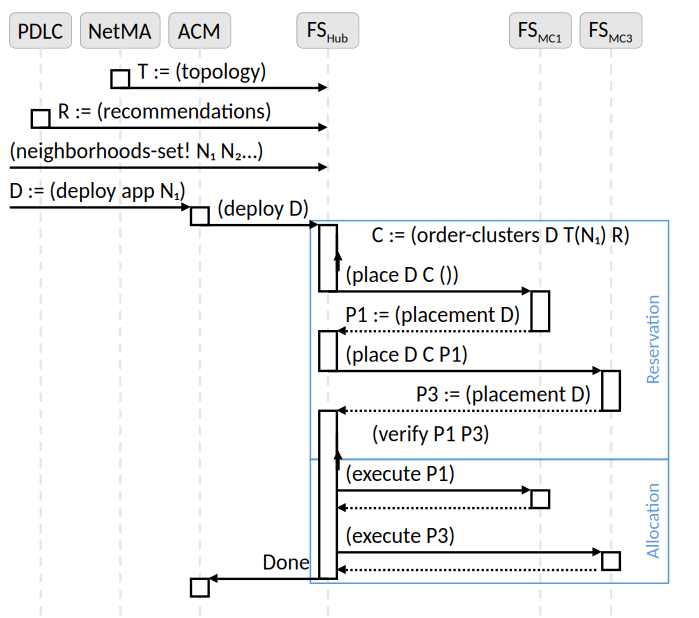}
  \caption{CODECO federated scheduling sequence: order clusters, solve L2
  problems sequentially, verify, then execute/rollback.}
  \label{fig:swm-fc-sequence}
\end{figure}

SWM-FC performs federated placement in three phases coordinated by the Hub instance; only the final phase allocates resources, as represented in Figure \ref{fig:swm-fc-sequence}:

\begin{enumerate}
  \item \textbf{Order clusters.} Compute an ordered sequence of clusters to visit
  within the neighborhood based on recommendations and inter-cluster link
  characteristics.
  \item \textbf{Solve cluster-by-cluster L2 placements.} Visit clusters in order.
  Each cluster solves an L2 problem using \emph{local} detailed information plus
  aggregate representations of the not-yet-visited clusters. The first visited
  cluster places as many workloads as possible (guided by node recommendations)
  and tentatively assigns remaining workloads to other clusters (as cluster-level
  nodes). The Hub integrates each partial result and forwards the updated problem
  to the next cluster.
  \item \textbf{Execute the placement.} Once all workloads are assigned and
  channels can be satisfied, SWM-FC creates pods and requests/installs the needed
  network services (intra- and inter-cluster). If execution fails in any target
  cluster, SWM-FC rolls back allocations across all affected clusters.
\end{enumerate}

A channel between workloads in two clusters is validated incrementally. When the
source workload is placed (e.g., cluster~A), SWM-FC minimizes the latency contribution within cluster~A (e.g., towards the cluster gateway/NED) to keep flexibility at the destination side. The final feasibility check occurs when the destination
workload is placed (cluster~B): total latency is computed as \textit{latency-in-A +
inter-cluster-link latency + latency-in-B}, and bandwidth feasibility (as well as any other requested resources) is checked
against the selected service class and available resources.

\subsubsection{Robustness: Reconsideration phase}
\label{subsubsec:swm-fc-robustness}
Because cluster internals are hidden outside each cluster, a first scheduling round may fail to place all workloads. To increase robustness, SWM-FC optionally
performs a \textit{reconsideration phase}: unplaced workloads are re-attempted by revisiting the ordered cluster list, relaxing placement constraints where allowed by policy (e.g., widening candidate sets while preserving hard QoS constraints). This second round is typically cheaper than the first because the remaining problem is smaller and most topology/context inputs are already cached at the participating instances.

\subsubsection{Workload migration}
\label{subsubsec:swm-fc-migration}

SWM-FC handles migration (in contrast to Kubernetes replication) using the same three-phase procedure as initial placement, triggered by changes in workload recommendations or infrastructure
conditions. Recommendations are  provided by PDLC, which may provide a recommendation for a "better" target node or cluster, in alignment with the target performance profiles set by the user in CAM, e.g., greenness, resilience, or others. During execution of a migration, SWM-FC reserves resources for the new placement before stopping the
old pod instance, enabling fast cut-over and reducing the risk of service disruption if the new instance cannot be created immediately.

\subsection{MDM-FC}
\label{mdm-fc}

The \textit{Metadata Management} component (MDM-FC) enables data-aware orchestration in CODECO by providing a unified, semantic view of data, applications, and infrastructure across CEI environments. Its primary objective is to support efficient and policy-compliant orchestration decisions by exposing data-related constraints and characteristics such as locality, freshness, portability, and compliance, to the control
plane.

MDM-FC integrates three complementary capabilities: (i) a semantic metadata catalog that captures cross-domain information about data, systems, and software; (ii) metadata discovery and enrichment, combining information collected from native systems with insights derived from multi-cluster operation; and (iii) data orchestration support, enabling CODECO to
account for data-related constraints when placing and adapting workloads (e.g., avoiding frequent access to remote datasets or enforcing compliance requirements that restrict data movement).

\paragraph{Semantic Metadata Model}
MDM represents metadata as a typed property graph, where entities and relationships can carry arbitrarily defined attributes. This model enables rich queries that derive application- and infrastructure-level metrics directly from the graph. Typical properties include data source and semantics, schema and data structure, application-specific characteristics (e.g., analysis or curation stage), and system-level attributes such as
compliance or observed performance.

The catalog is populated continuously using both push and pull mechanisms. Metadata may be ingested directly from Kubernetes APIs, external data sources, or user input propagated through ACM-FC and Prometheus metrics. MDM maintains this information as a shared abstraction layer, allowing metadata originating from multiple clusters and domains to be reasoned about
in a uniform way.

To decouple metadata ingestion from storage and querying, MDM relies on an event-driven architecture based on Apache Kafka. Metadata changes are represented as events (insert,
update, delete) affecting entities or relationships. These events are logged and selectively
materialized into the metadata graph to satisfy the needs of other CODECO components, notably
PDLC-FC.

\paragraph{Entities, Relationships, and Extensibility.}
Example entities in a CODECO/Kubernetes environment include \textit{pods}, \textit{namespaces}, \textit{worker nodes}, as well as data-related objects such as \textit{datastores} or \textit{files}. Relationships are typed, directional, and connect exactly two entities (e.g.,
a \textit{contains} relationship linking a namespace to its pods). New entity and relationship types can be defined using a flexible JSON Schema DSL~\cite{pathfinder}. This extensibility allows MDM to integrate metadata from arbitrary native systems without requiring changes to the core data model.

\subsubsection{Components}
\begin{figure}[t!]
  \centering
  \includegraphics[width=\linewidth]{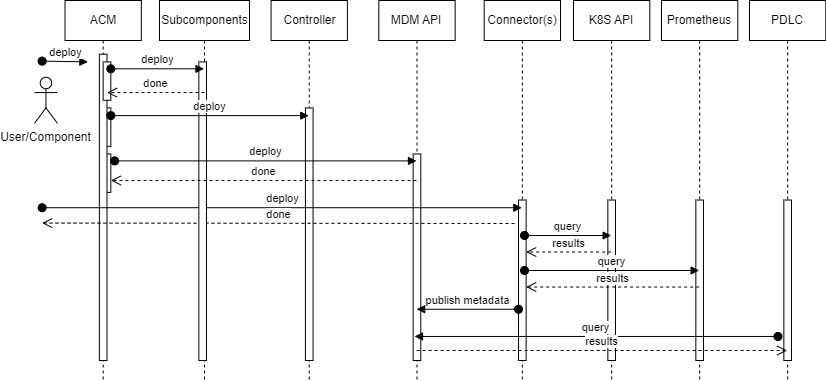}
  \caption{MDM components and interaction within CODECO.}
  \label{fig:mdm-components-flow}
\end{figure}
MDM-FC is implemented as three loosely coupled components (Figure~\ref{fig:mdm-components-flow}):

\begin{itemize}
    \item \textbf{Graph Database}. The graph database stores the materialized metadata graph and provides a global view over entities and relationships collected from distributed sources. In the current implementation,
Neo4j is used, but any Cypher-capable graph database can be adopted. Direct graph queries are possible during development and exploration, but production access is mediated through the MDM API to preserve abstraction and portability.
    \item \textbf{MDM API}. The MDM API exposes OpenAPI-compliant REST endpoints that allow metadata to be ingested, queried, and consumed by other CODECO components. The API includes endpoints for publishing metadata events, accessing the event stream, querying the metadata graph, and retrieving
domain-specific views required by PDLC-FC (e.g., pod-level portability, compliance, or freshness metrics). Kafka is used internally to decouple API ingestion from graph storage, ensuring scalability and resilience.
    \item \textbf{Connectors}. Connectors are responsible for collecting metadata from native systems and publishing it to MDM as structured events. Each connector manages the lifecycle of the entities it inserts, ensuring isolation and preventing cross-connector interference. Entities are assigned deterministic global identifiers derived from stable attributes (e.g., cluster ID, namespace, pod name), enabling consistent  identification across clusters.
    \end{itemize}

Out of the box, CODECO provides three connectors commonly used by PDLC-FC:
(i) a \textit{Kubernetes connector}, which reflects Kubernetes objects and relationships into the metadata graph and provides a proxy for portability through pod restart counts;
(ii) a \textit{Prometheus connector}, which enriches entities with application-level metrics such as freshness; and (iii) a \textit{Compliance connector}, which integrates cluster compliance scores (e.g., via Kubescape) to support policy-aware placement decisions.

\subsubsection{Federated Operation}
MDM-FC is designed to operate efficiently in federated environments without requiring tight coupling between clusters. While a fully symmetric federation with global metadata replication is conceptually possible, CODECO adopts a neighborhood-scoped deployment model that aligns with its bounded coordination principle.

\paragraph{Single neighborhood per cluster}
Within a neighborhood, a single cluster acts as the MDM hub, hosting the graph database and MDM API for that neighborhood. Connectors deployed on Managed Clusters report metadata to the hub instance, and PDLC-FC instances query the hub to retrieve data-related metrics relevant to the applications deployed in the neighborhood. This avoids maintaining a global metadata
view when it is not required for orchestration.

If a Managed Cluster leaves a neighborhood, its metadata simply stops being updated. If the MDM hub becomes unavailable, ACM-FC can redeploy the MDM instance on another cluster and reconfigure connectors accordingly. Although this failover scenario is supported by design, it is not considered part of the standard operational workflow.

\paragraph{Supporting Multiple neighborhoods}
\begin{figure}[t!]
  \centering
  \includegraphics[width=\linewidth]{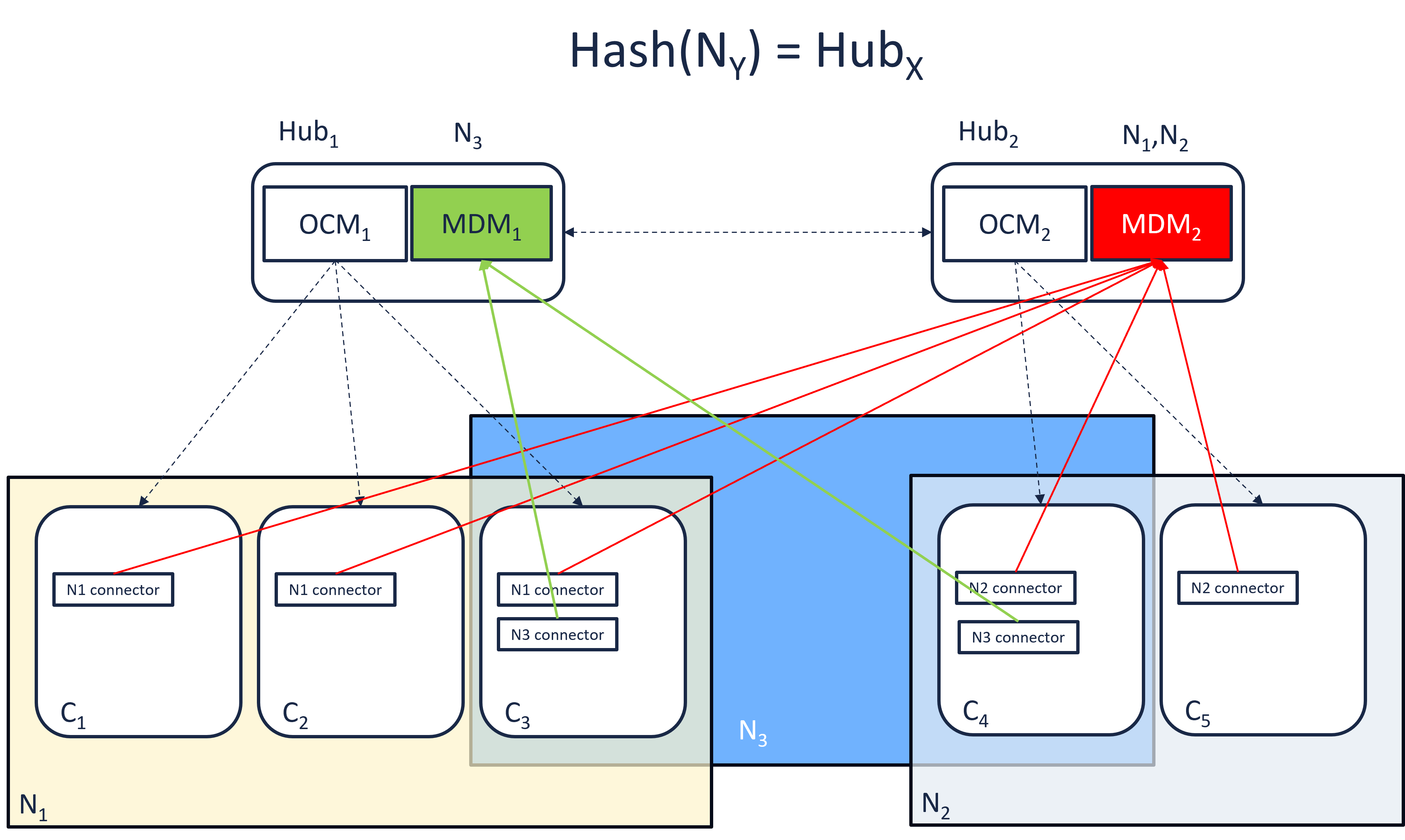}
  \caption{MDM supporting multiple neighborhoods per instance.}
  \label{fig:mdm-multi-neighborhood}
\end{figure}
When clusters participate in multiple neighborhoods, MDM-FC must avoid deploying a separate metadata stack for each neighborhood on the same hub. Figure~\ref{fig:mdm-multi-neighborhood} illustrates how a limited number of MDM instances hosted on hub clusters can jointly support multiple, potentially overlapping neighborhoods by consistently mapping neighborhood
identifiers to MDM instances.

Connectors can be configured to report to multiple MDM instances, allowing a single connector deployment per cluster to serve several neighborhoods and reducing overhead. For large-scale deployments, hub clusters may maintain a distributed hash table (DHT)~\cite{DHT} to manage neighborhood-to-MDM mappings efficiently.

\section{Workflow Examples}
\label{sec: workflow}

This section illustrates the end-to-end operation of CODECO through three representative workflows: (i) installation and deployment of the CODECO framework across federated clusters, (ii) deployment of an application across a Cloud–Edge Infrastructure (CEI), and (iii) runtime management and adaptation of applications. The workflows reflect the interactions described in the previous sections and emphasize how orchestration, monitoring, and learning components collaborate at different lifecycle stages.

\subsection{CODECO Installation and Deployment}
\begin{figure}[t!]
  \centering
  \includegraphics[width=0.9\columnwidth]{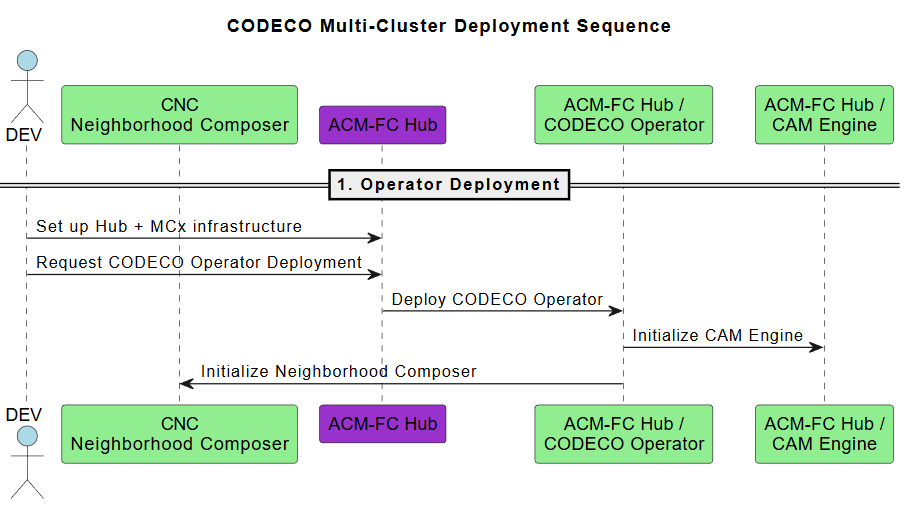}
  \caption{Sequence diagram illustrating the installation and deployment of CODECO across federated clusters with one Hub and multiple Managed Clusters.}
  \label{fig:codeco-seq}
\end{figure}

\begin{figure*}[t!]
  \centering
  \includegraphics[width=0.9\textwidth]{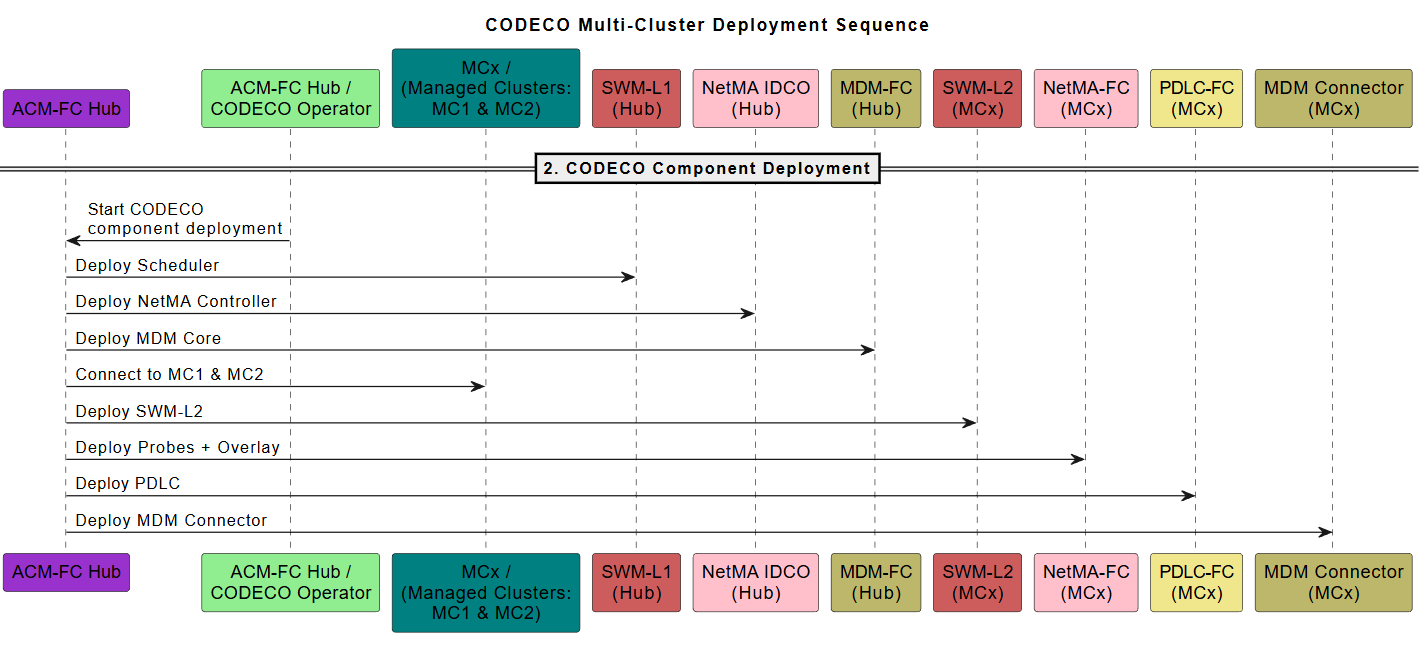}
  \caption{Sequence diagram illustrating the installation and deployment of CODECO across federated clusters with one Hub and multiple Managed Clusters.}
  \label{fig:codeco-components-deploy}
\end{figure*}

\begin{figure}[t!]
  \centering
  \includegraphics[width=0.9\columnwidth]{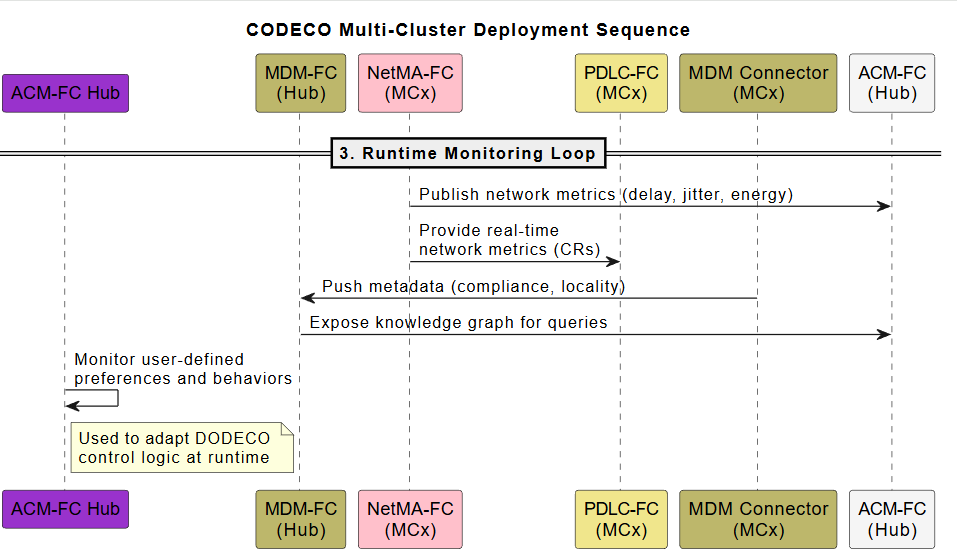}
  \caption{Sequence diagram illustrating the installation and deployment of CODECO across federated clusters with one Hub and multiple Managed Clusters.}
  \label{fig:runtime}
\end{figure}

To install the \textit{CODECO OSS Federated Toolkit}, a developer (DEV) retrieves the software from the official CODECO repository and follows the installation guidelines. A prerequisite is an operational OCM environment comprising at least one Hub cluster and a set of Managed Clusters (MCs). The complete process unfolds in three phases, as summarized in Figures~\ref{fig:codeco-seq}–\ref{fig:runtime}.

\textbf{Phase~1: Operator deployment.} The DEV first prepares the infrastructure by registering the Managed Clusters with the Hub via OCM. Using the ACM-FC interface, the DEV triggers the deployment of the CODECO Operator on the Hub. This step initializes the core control-plane services required for federated orchestration, including the CAM Engine for parsing and validating application models and the CODECO CNC which enables semantic grouping of clusters.

\textbf{Phase~2: Component deployment.} Once the operator is active, ACM-FC orchestrates the deployment of CODECO components across the Hub and the MCs. On the Hub, the following components are
instantiated:
\begin{itemize}
    \item \textbf{SWM-FC (Hub instance)}, coordinating federated scheduling and
    workload migration,
    \item \textbf{NetMA IDCO}, managing inter-cluster secure connectivity and
    overlay control,
    \item \textbf{MDM-FC}, maintaining the metadata graph for applications,
    infrastructure, and compliance aspects.
\end{itemize}

On each Managed Cluster, ACM-FC deploys:
\begin{itemize}
    \item \textbf{SWM-FC (MC instance)}, responsible for local workload placement
    and execution,
    \item \textbf{NetMA-FC}, providing network probing, overlay support, and
    metric exposure,
    \item \textbf{PDLC-FC}, executing context-aware cost estimation and
    decentralized learning,
    \item \textbf{MDM connectors}, feeding local metadata into the neighborhood
    MDM instance.
\end{itemize}

After deployment, continuous monitoring loops are activated. NetMA-FC starts
collecting network-level metrics (e.g., latency, jitter, packet loss, energy),
MDM connectors populate and update the metadata graph, and ACM observes user
preferences and application intents. These data streams form the baseline for
subsequent orchestration and adaptation decisions.

\subsection{Deploying an Application across a Federated CEI Environment}
The deployment workflow proceeds as follows:
\begin{enumerate}
  \item Submission and validation of the CAM at ACM-FC.
  \item Neighborhood selection via the CNC.
  \item Initialization of PDLC-FC agents within the neighborhood.
  \item Federated placement and channel assignment by SWM-FC.
  \item Network provisioning and secure connectivity setup by NetMA-FC.
\end{enumerate}

\subsection{Management of Applications during their Runtime}

After deployment, CODECO enters the runtime management phase, during which it continuously observes application and infrastructure behaviour and adapts the deployment when necessary as illustrated in Figure~\ref{fig:runtime}.

Consider an application group comprising two microservices, $a_1$ and $a_2$,
deployed on Managed Clusters MC1 and MC2, respectively, under the same Hub. At runtime, monitoring data from NetMA-FC and ACM, together with metadata updates
from MDM, are continuously fed into PDLC-FC. If a triggering condition arises (e.g., performance degradation, resource contention, energy constraints, or a
change in application intent), ACM-FC signals the need for re-evaluation.

PDLC-FC responds by updating its cost estimates and learning-based stability assessments and publishes revised recommendations via CRs. SWM-FC observes these updates and initiates a new placement computation, which may result in workload migration within or across clusters, provided that all constraints can be met.

Once a new assignment is decided, SWM-FC updates the relevant CRs
(\texttt{AssignmentPlan}, \texttt{ApplicationGroup}), triggering NetMA-FC to reconfigure network paths if needed. Throughout this process, coordination between components is achieved exclusively through Kubernetes CRs, avoiding
tight coupling or direct control-plane dependencies.

Finally, ACM-FC aggregates the current system and application state and exposes it to the DEV through dashboards or APIs, enabling transparency and informed
human oversight while CODECO autonomously manages the application lifecycle.

\section{CODEF: Experimentation and Software-defined Testbed}
\label{sec:experimentation}

The evaluation of decentralized orchestration mechanisms for federated CEI environments requires a reproducible, automated, and scalable experimentation strategy capable of capturing the  interactions between compute, data, and network resources. In contrast to single-cluster experimentation, federated scenarios introduce additional challenges related to geographic distribution, heterogeneous infrastructures, multi-network connectivity, and dynamic operational conditions.

This section presents the experimentation methodology adopted in CODECO and which is embodied in the \textit{CODECO Experimentation Framework (CODEF)}~\cite{codef2024} software-defined Edge-Cloud Emulation tool. CODEF provides an end-to-end, software-defined experimentation environment that streamlines infrastructure provisioning, platform configuration, experiment execution, and result collection in a one-touch, automated manner.

Furthermore, we describe how CODEF is being extended to support large-scale, geographically distributed federated testbeds, enabling realistic validation of CODECO’s decentralized orchestration mechanisms.

\subsection{CODEF as Individual Emulator}

\begin{figure}[t]
  \centering
  \includegraphics[width=\columnwidth]{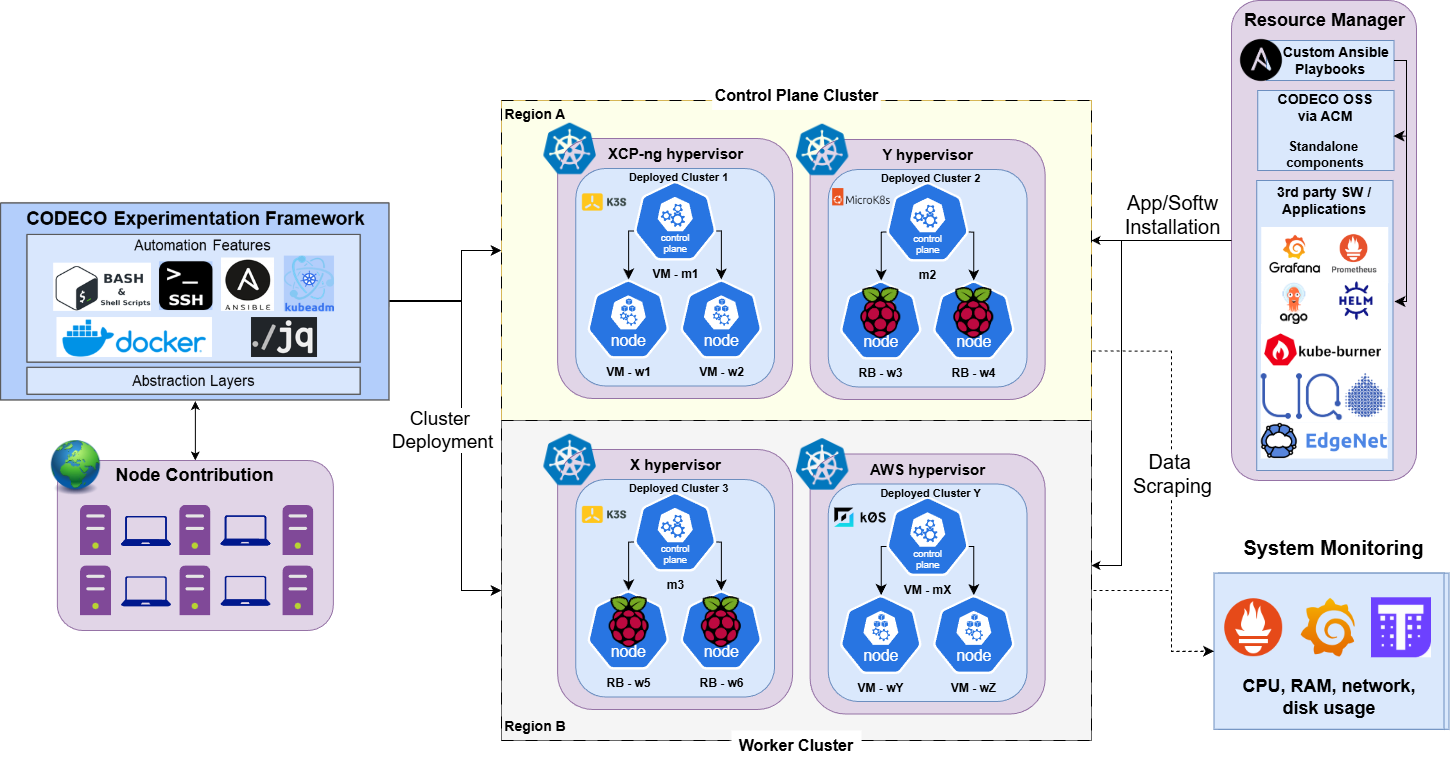}
  \caption{CODEF end-to-end experimentation workflow.}
  \label{fig:codef-workflow}
\end{figure}

CODEF provides an automated end-to-end workflow for designing, deploying, and evaluating experiments in federated CEI environments (Figure~\ref{fig:codef-workflow}). It is implemented as a microservice-based framework with minimal prerequisites (e.g., Docker and standard CLI tools), and introduces a clear separation of concerns across the experimentation lifecycle.

Specifically, CODEF distinguishes between:
\begin{itemize}
    \item \textbf{Infrastructure management}, responsible for provisioning compute resources across edge, cloud, and virtualized environments;
    \item \textbf{Resource management}, which installs operating systems, Kubernetes distributions, Container Network Interface plugins (CNIs), and required software stacks;
    \item \textbf{Experiment execution}, covering parameterized workload deployment, orchestration policies, iterations, and fault, traffic or load injection;
    \item \textbf{Result processing}, including metric collection, aggregation, generation of reproducible outputs and reporting.
\end{itemize}

All CODEF components are containerized microservices, allowing the framework to be easily extended with new technologies, orchestration tools, or evaluation capabilities. This design supports continuous evolution of the experimentation platform while maintaining reproducibility.

A key capability of CODEF is its support for heterogeneous infrastructures. \textbf{Infrastructure Managers} abstract the provisioning of nodes across edge–cloud and virtualized environments (e.g., XCP-ng\footnote{https://xcp-ng.org/}, VirtualBox\footnote{https://www.virtualbox.org/}, AWS\footnote{https://aws.amazon.com/}), and allow integration with external research testbeds such as CloudLab\footnote{https://www.cloudlab.us/} and FABRIC\footnote{https://portal.fabric-testbed.net/}. 

\textbf{Resource Managers} represent each node as a containerized entity and deploy operating systems, Kubernetes distributions, CODECO components, and experimental workloads using Ansible playbooks.

At the orchestration level, a declarative \textbf{Experiment Controller} unifies all experiment parameters such as infrastructure topology, Kubernetes configuration, networking fabric, workload definitions, software versions, metrics, and output formats into a single, versioned specification. This enables fully reproducible, one-touch experiment execution.

For federated scenarios, CODEF integrates established multi-cluster and multi-networking solutions, including OCM, Submariner\footnote{https://submariner.io/} and Liqo\footnote{https://liqo.io/}. This enables experimentation across a wide range of federation models, from hub-and-spoke and neighborhood-based deployments to peer-to-peer cluster meshes. As a result, clusters and individual nodes may be geographically distributed while remaining coordinated within a single federated orchestration domain.

\subsection{CODEF as a Collaborative, Multi-tenant Federated Testbed}
In addition to its emulation role, CODEF is being extended in CODECO to be used as an open federated testbed that relies on node contributions by different entities across the Internet to emulate a true collaborative CEI multi-tenant, federated environment, based on the principles of the now extinct EdgeNet\footnote{https://www.edge-net.org/}. This collaborative operational environment allows users, organizations, and external testbeds to contribute compute resources (virtual machines or bare-metal nodes) into a shared federated environment\footnote{In 2025, the current testbed interconnects CODEF based on node contributions in Greece and Germany. The opening to the public is expected in 2026. The interface to the federated CODEF testbed is available at: \url{https://boss.swn.uom.gr/}}. 

Inspired by large-scale research infrastructures, this capability enables crowd-sourced, geographically dispersed multi-cluster deployments that more accurately reflect real-world CEI conditions. 

\subsection{Prototyping and Evaluation Workflows}
Evaluating decentralized orchestration in federated environments requires well-defined prototyping and evaluation workflows that can accommodate heterogeneous hardware, diverse networking fabrics, and dynamic CEI-scale conditions. These workflows must remain platform-agnostic, ensuring that experiments can be executed consistently across different Kubernetes distributions, virtualization technologies, and edge–cloud deployments.

A typical workflow begins with the declarative specification of the experimental scenario, including cluster composition, inter-cluster connectivity, and orchestration policies. Once clusters are provisioned and federated, application workloads are deployed to evaluate operational challenges such as variable latency, intermittent connectivity, energy constraints, and resource contention.

During execution, cross-layer observability mechanisms collect compute-, network-, and data-plane metrics using a combination of passive monitoring, active probing, and integrated telemetry. These measurements enable systematic evaluation of scheduling decisions, application behaviour, and adaptation or migration strategies driven by components such as PDLC, NetMA, and SWM. The proposed workflows align with ongoing standardization efforts in bodies such as ETSI and IETF, which aim to define interoperable mechanisms for multi-cluster orchestration, intent-driven management, and federated telemetry.

Within this context, CODEF provides the automation, reproducibility, and scalability required to operationalize such workflows, bridging the gap between conceptual orchestration models and realistic CEI experimentation.

\subsection{Reproducibility and Open-source Toolkit}
\label{sec:oss-toolk}

Reproducibility is a fundamental requirement for experimental validation, particularly in federated environments where infrastructure diversity, geographic distribution, and operational variability can significantly affect experimental outcomes. Open-source toolkits, such as those developed within the CODECO ecosystem, play a crucial role in enabling transparent experimentation, innovation, and extensibility.

By adopting a software-defined approach, CODEF abstracts platform heterogeneity and provides stable, declarative interfaces for deploying clusters, configuring federations, and executing orchestration experiments at scale. This enables researchers and practitioners to replicate experiments, validate results, and extend the framework with new orchestration strategies, metrics, or infrastructure technologies. As a result, CODEF establishes a robust foundation for reproducible, scalable, and collaborative research on decentralized orchestration across the CEI continuum.

\section{Research Challenges and Future Directions} 
\label{sec:future}

Despite the advances introduced by CODECO, federated orchestration across the CEI continuum raises a number of open research challenges. These challenges are not specific to CODECO but are representative of a broader class of decentralized, multi-provider orchestration systems operating under heterogeneous infrastructure, administrative, and regulatory constraints \cite{ullah2023a, hong2019a, baldoni2021a}.

\subsection{Scheduling Challenges in Federated Container Orchestration}

A fundamental challenge lies in balancing optimality, scalability, and privacy in federated scheduling. In multi-cluster environments, global optimization typically requires access to fine-grained infrastructure and workload state, which is often unavailable or undesirable due to privacy, ownership, or scalability constraints \cite{larsson2020b, cannarella2022a}. CODECO addresses this tension through neighborhood-scoped scheduling and aggregated cluster representations; however, defining optimal aggregation strategies that preserve scheduling quality while minimizing information leakage remains an open problem.

Another challenge concerns \textit{cross-layer coupling}. While CODECO jointly considers compute, network, and data aspects, accurately capturing their interdependencies particularly under dynamic conditions such as mobility, intermittent connectivity, and workload churn—remains difficult. Prior work has highlighted similar limitations in fog and edge orchestration \cite{hong2019a, ullah2023a}. Future work may explore tighter integration of predictive models, uncertainty-aware scheduling, and stochastic optimization techniques \cite{zheng2022a, xiong2020a}.

\subsection{Stability, Oscillation, and AI-assisted Control}
AI-assisted orchestration introduces the risk of oscillatory behavior when learning-based components react too aggressively to transient fluctuations in resource availability or workload demand \cite{bu2010a, browne2019a}. CODECO mitigates this risk by separating recommendation from enforcement and constraining learning components to medium timescales. Nonetheless, formal stability guarantees for hybrid AI–rule-based orchestration systems remain largely unexplored and represent an important research direction, particularly for safety-critical CEI deployments \cite{zhang-a, bellavista2021}.

\subsection{Multi-hub and Cross-federation Operation}
The current CODECO design assumes a single Hub cluster per federation. Supporting multiple hubs each governing its own administrative domain—raises challenges related to namespace collisions, trust management, policy reconciliation, and cross-hub neighborhood composition \cite{larsson2020b, enel-a}. Extending CODECO toward hierarchical or peer-to-peer hub federation aligns with emerging visions of decentralized CEI governance and remains a natural next step.

\section{Related Work and Comparative Analysis} \label{sec:relatedwork}
Kubernetes multi-cluster and edge orchestration has evolved into a rich ecosystem of
projects and research prototypes. Having described CODECO’s design and operation, this section now explains the contributions of CODECO in terms of (i) Kubernetes federation and multi-cluster control planes, (ii) edge-centric Kubernetes
distributions and frameworks, (iii) inter-cluster connectivity and service networking,
(iv) learning-assisted schedulers for federated clusters, and (v) data/metadata-aware
orchestration.

\subsection{Federation and Multi-cluster Control Planes}
Early federation efforts, such as Kubernetes Federation (KubeFed\footnote{https://github.com/kubernetes-retired/kubefed}), focused primarily on replicating resources and enabling cross-cluster deployment primitives \cite{larsson2020b}. These approaches offer limited support for cross-layer optimization and do not explicitly model network and data constraints as first-class scheduling inputs \cite{casalicchio2019a}.

OCM generalizes this direction with a production-grade hub-and-spoke architecture for cluster registration, policy propagation, and manifest distribution \cite{ocm_project}. OCM provides the governance substrate required in many enterprise and multi-tenant settings, but it is not a workload optimizer by itself. CODECO builds on OCM as the governance layer and adds bounded federation via application neighborhoods, together with a co-orchestration loop integrating compute, network, and data observability into placement and adaptation decisions \cite{sofia2024framework}.

Karmada extends Kubernetes with multi-cluster scheduling, propagation policies, and resource placement across clusters, providing a practical control plane for multi-cloud operation \cite{karmada_project}. Its design largely treats networking and data constraints as external signals. CODECO complements such systems with an explicit separation between recommendation (PDLC-FC) and enforcement (SWM-FC), and with network- and data-aware inputs (NetMA-FC and MDM-FC) exposed through Kubernetes resources and Prometheus.

\subsection{Federated Scheduling Approaches}
PHARE targets placement of multi-component applications across federated edge clusters and explicitly addresses limited local resources and cross-cluster composition under network variability \cite{galantino2024phare}. While PHARE demonstrates the importance of federation-aware scheduling, it does not provide an integrated stack for privacy-preserving learning, network capability exposure, and data governance under a unified Kubernetes-native control interface.

CODECO extends this line of work by combining (i) neighborhood-scoped federation, (ii) privacy-preserving decentralized learning (PDLC-FC), (iii) programmable secure inter-cluster overlays and ALTO-style cost maps (NetMA-FC), and (iv) metadata-graph-based observability (MDM-FC), enabling scheduling decisions that are jointly data–compute–network aware.

\subsection{Edge-centric Kubernetes Frameworks}
KubeEdge\footnote{https://kubeedge.io/} extends Kubernetes with edge autonomy, device management, and intermittent-connectivity tolerance via a cloud/edge split control plane \cite{mueller-a, gogouvitis2020a}. Its strengths lie in edge execution and device integration, while cross-provider federation and multi-cluster optimization are not its primary focus. CODECO is complementary, assuming Kubernetes at the edge while focusing on federated orchestration across administrative domains.

OpenYurt\footnote{https://openyurt.io/} similarly emphasizes edge autonomy and minimal deviation from upstream Kubernetes \cite{openyurt_project}. In contrast, CODECO elevates federation as a first-class concern, combining OCM-based governance with learning-assisted recommendations and explicit network and data constraints to support placement and migration across clusters.

\subsection{Inter-cluster Connectivity and Service Networking}
Submariner provides pragmatic inter-cluster connectivity and service discovery \cite{submariner_project}. While effective at enabling communication, it does not expose a unified, multi-metric network view to schedulers. CODECO builds on such primitives but extends them by making network metrics (latency, bandwidth, loss, energy) observable and actionable via NetMA-FC and ALTO-style cost maps \cite{alimi2014a, camaz2020a}.

Cilium Cluster Mesh offers scalable multi-cluster networking and policy enforcement using eBPF \cite{qi-b}. CODECO differs by tightly coupling monitored network costs and service classes with orchestration decisions, integrating them into PDLC-FC recommendations and SWM-FC enforcement.

\subsection{Learning-assisted Scheduling for Federated Kubernetes}
Learning-based schedulers have been proposed to improve placement and adaptation in distributed and federated environments \cite{huang-a, zheng2022a, senjab2023survey}. These approaches often rely on centralized telemetry or global training. CODECO introduces a decentralized, privacy-preserving learning pipeline that keeps raw telemetry local, exchanges only protected learned artifacts, and separates recommendation from enforcement to preserve stability and auditability \cite{sofia2024framework}.

\subsection{Data/Metadata-aware Orchestration and Compliance}
Data-aware orchestration is often handled through ad-hoc affinity rules or external policy engines. Research on enterprise metadata graphs and data fabrics has shown the value of richer metadata integration \cite{pathfinder, ghiran-a}. CODECO contributes an explicit metadata management plane (MDM-FC) that exposes data locality, freshness, and compliance as queryable orchestration inputs, enabling placement decisions that remain policy-compliant across federated clusters.

\subsection{Comparison with Federated Kubernetes Approaches}
\label{subsec:comparison}

\begin{table*}[ht!]
\centering
\caption{Comparison of CODECO with representative federated Kubernetes approaches.}
\label{tab:comparison-codeco}
\scriptsize
\setlength{\tabcolsep}{4pt}
\renewcommand{\arraystretch}{1.05}
\begin{tabular}{p{3.1cm} p{3.2cm} p{2.7cm} p{2.7cm} p{2.7cm}}
\hline
\textbf{Dimension} &
\textbf{CODECO} &
\textbf{Liqo} &
\textbf{PHARE} &
\textbf{Karmada} \\
\hline

Primary focus &
Cross-layer co-orchestration &
Resource sharing, offloading &
Federated edge placement &
Multi-cluster propagation \\

Federation model &
Bounded neighborhoods &
Peer-to-peer virtual clusters &
Federated edge clusters &
Hub-and-spoke \\

Governance &
Hybrid (central policy, local exec.) &
Decentralized peering &
Centralized per deployment &
Centralized hub \\
\hline

Scheduling scope &
Neighborhood-scoped &
Local + remote offloading &
Cross-cluster &
Cross-cluster \\

Scheduling strategy &
Deterministic + AI guidance &
Scheduler extensions &
Heuristic / optimization &
Policy-driven \\

AI / learning &
Decentralized, privacy-preserving &
None &
None &
None \\
\hline

Network awareness &
First-class (metrics, cost maps) &
Basic connectivity &
Limited / implicit &
Limited (policy hints) \\

Inter-cluster networking &
Secure overlays, SDN-assisted &
Gateway-based &
External / assumed &
External (e.g., Submariner) \\
\hline

Data / metadata awareness &
Explicit metadata graph &
None &
Limited &
None \\

Privacy and sovereignty &
Local telemetry, protected exchange &
Trust-based sharing &
Not addressed &
Not addressed \\
\hline

Edge autonomy &
Designed for intermittency &
Partial &
Limited &
Limited \\

Multi-provider support &
Explicitly targeted &
Implicit &
Research-oriented &
Enterprise / cloud \\

Runtime adaptation &
Proactive + reactive &
Reactive &
Reactive &
Reactive \\

Kubernetes integration &
Native (CRDs, OCM-based) &
Native &
Prototype &
Native \\
\hline
\end{tabular}
\end{table*}

Table~\ref{tab:comparison-codeco} compares CODECO with representative federated and multi-cluster Kubernetes approaches, namely Liqo, PHARE,
and Karmada. The comparison highlights architectural scope, federation models, scheduling strategies, network and data awareness, and support
for AI-assisted orchestration. CODECO differentiates itself by combining bounded federation via application neighborhoods with cross-layer observability and privacy-preserving decentralized learning, enabling cognitive co-orchestration across the CEI continuum.

\section{Conclusions} \label{sec:conclusions}

This paper presents CODECO, a cognitive and decentralized orchestration framework
that extends Kubernetes to support federated operation across the
CEI continuum. CODECO addresses the limitations of
cloud-centric orchestration by introducing a data-compute-network
co-orchestration model, bounded federation through application neighborhoods,
and AI-assisted decision support that remains compatible with Kubernetes-native
control and governance mechanisms.

The core contribution of CODECO lies in its architectural separation of concerns.
Centralized governance is preserved through OCM-based hub control and policy
enforcement, while execution, monitoring, and learning are decentralized and
scoped to application-relevant neighborhoods. This hybrid model enables
scalability, privacy preservation, and resilience under intermittent
connectivity, without resorting to full global state sharing or tightly coupled
control loops.

CODECO further demonstrates how cross-layer observability can be elevated to a
first-class orchestration input. By integrating network awareness (NetMA-FC),
data and compliance metadata (MDM-FC), and privacy-preserving decentralized
learning (PDLC-FC), CODECO enables informed placement, migration, and adaptation
decisions that go beyond compute-centric scheduling. The explicit separation
between learning-based recommendation and deterministic enforcement mitigates
the risk of instability and oscillations often associated with AI-driven
orchestration.

To support reproducible evaluation, the paper introduced CODEF, a
software-defined experimentation framework that enables automated deployment,
execution, and analysis of federated CEI scenarios. CODEF complements CODECO by
providing a practical foundation for large-scale experimentation across
heterogeneous infrastructures and geographically distributed testbeds.

While CODECO already demonstrates a comprehensive and extensible design,
important research challenges remain. These include optimal aggregation
strategies for privacy-preserving federated scheduling, formal stability
guarantees for hybrid AI-assisted control loops, and support for multi-hub and
cross-federation operation. Addressing these challenges will be essential for
deploying cognitive orchestration frameworks at Internet scale.

\section*{Acknowledgments}
This work has been funded by The European Commission in the context of the Horizon Europe CODECO project under grant number 101092696, and by the Swiss State Secretariat for Education, Research and Innovation (SERI) under contract number 23.00028.

We thank all members of the CODECO consortium for their valuable contributions to CODECO deliverables D8, D9, and D11. In addition, we thank in particular the following colleagues: Tina Samizadeh, fortiss, for her contributions to the CODECO standardization and experimentation; Dorine Matzakou, Netcompany, for her invaluable contributions to the CODECO learning hub and the CODECO exploitation plan; Andries Stam, Almende, for his leadership in the CODECO sustainability and asset exploitation plan; Ignacio Prusiel, Eviden, for his contributions to the initial development of the CODECO observability layer; David Remon and Marco Jahn for the invaluable contributions to the development of the CODECO open-source community, and the Eclipse KUDECO project.

\bibliographystyle{IEEEtran}
\bibliography{codecoreferences}

@misc{raft,
	title        = {{In Search of an Understandable Consensus Algorithm}},
	author       = {{Ongaro, Diego and Ousterhout, John}},
	year         = 2014
}

@article{galantino2024phare,
	title        = {{Scheduling Multi-Component Applications Across Federated Edge Clusters With {Phare}}},
	author       = {{Galantino, Stefano and Filaferro, Gabriele and Nicosia, Luca}},
	year         = 2024,
	journal      = {{IEEE Open Journal of the Communications Society}},
	volume       = 5,
	pages        = {1814--1826},
	doi          = {10.1109/OJCOMS.2024.3377917}
}

@misc{karmada_project,
	title        = {{{Karmada}: Kubernetes multi-cluster management (project)}},
	author       = {{Karmada Project}},
	note         = {Accessed: 2026-01-09},
	howpublished = {\url{https://github.com/karmada-io/karmada}}
}

@misc{ocm_project,
	title        = {{{Open Cluster Management}: Multi-cluster lifecycle and policy management for Kubernetes (project)}},
	author       = {{Open Cluster Management Authors}},
	note         = {Accessed: 2026-01-09},
	howpublished = {\url{https://github.com/open-cluster-management-io}}
}

@misc{submariner_project,
	title        = {{{Submariner}: Cross-cluster connectivity for Kubernetes (project)}},
	author       = {{Submariner Project}},
	note         = {Accessed: 2026-01-09},
	howpublished = {\url{https://github.com/submariner-io/submariner}}
}

@misc{openyurt_project,
	title        = {{{OpenYurt}: Extending Kubernetes to edge computing (project)}},
	author       = {{OpenYurt Project}},
	note         = {Accessed: 2026-01-09},
	howpublished = {\url{https://github.com/openyurtio/openyurt}}
}

@inproceedings{feeney2001,
	title        = {{Investigating the energy consumption of a wireless network interface in an ad hoc networking environment}},
	author       = {{Feeney, L.M. and Nilsson, M.}},
	year         = 2001,
	booktitle    = {Proceedings IEEE INFOCOM 2001. Conference on Computer Communications. Twentieth Annual Joint Conference of the IEEE Computer and Communications Society (Cat. No.01CH37213)},
	volume       = 3,
	number       = {},
	pages        = {1548--1557 vol.3},
	doi          = {10.1109/INFCOM.2001.916651}
}

@article{sofia2024framework,
  author       = {Sofia, Rute C. and Salomon, Josh and Ferlin-Reiter, Simone and Garc{\'e}s-Erice, Luis and Urbanetz, Peter and Mueller, Harald and Touma, Rizkallah and Espinosa, Alejandro and Contreras, Luis M. and Theodorou, Vasileios and others},
  title        = {{A Framework for Cognitive, Decentralized Container Orchestration}},
  journal = {{IEEE Access}},
  year         = {2024},
  volume       = {12},
  pages        = {79978--80008},
  doi          = {10.1109/ACCESS.2024.3377917},
}

@online{alimi2014a,
  author       = {Alimi, Reinaldo and Penno, Reinaldo and Yang, Y. Richard and Kiesel, Sebastian and Previdi, Stefano and Roome, Will and Shalunov, Stanislav and Woundy, Rich},
  title        = {{Application-Layer Traffic Optimization (ALTO) Protocol}},
  institution  = {IETF},
  type         = {RFC},
  number       = {7285},
  year         = {2014-09},
  url          = {https://www.rfc-editor.org/rfc/rfc7285},
}

@online{brader1997a,
  author       = {Bradner, Scott},
  title        = {{Key Words for Use in RFCs to Indicate Requirement Levels}},
  institution  = {IETF},
  type         = {RFC},
  number       = {2119},
  year         = {1997-03},
  url          = {https://www.rfc-editor.org/rfc/rfc2119},
}

@inproceedings{abowd-a,
  author       = {Abowd, Gregory D. and Dey, Anind K. and Brown, Peter J. and Davies, Nigel and Smith, Mark and Steggles, Pete},
  title        = {{Towards a Better Understanding of Context and Context-Awareness}},
  booktitle    = {{Proceedings of the 1st International Symposium on Handheld and Ubiquitous Computing}},
  publisher    = {Springer},
  year         = {1999},
  pages        = {304--307},
  doi          = {10.1007/3-540-48157-5_29},
}

@inproceedings{liu2011a,
  author       = {Liu, Wei and Li, Xiaodong and Huang, Deyi},
  title        = {{A Survey on Context Awareness}},
  booktitle    = {{2011 International Conference on Computer Science and Service System (CSSS)}},
  publisher    = {IEEE},
  year         = {2011},
  pages        = {144--147},
}

@incollection{sofia2023role,
  author       = {Sofia, Rute C.},
  title        = {{The Role of Socially Aware Networking in Supporting 6G IoT}},
  booktitle    = {{6G Visions for a Sustainable and People-Centric Future}},
  publisher    = {River Publishers},
  year         = {2023},
  pages        = {55--78},
}

@misc{cannarella2022a,
  author       = {Cannarella, Alessandro},
  title        = {{Multi-Tenant Federated Approach to Resource Brokering between Kubernetes Clusters}},
  institution  = {Politecnico di Torino},
  type         = {PhD diss.},
  year         = {2022},
}

@article{ullah2023a,
  author       = {Ullah, Amjad and Kiss, Tamas and Kov{\'a}cs, J{\'o}zsef and Tusa, Francesco and Deslauriers, James and Dagdeviren, Huseyin and Hamzeh, Hamed},
  title        = {{Orchestration in the Cloud-to-Things Compute Continuum: Taxonomy, Survey and Future Directions}},
  journal = {{Journal of Cloud Computing}},
  year         = {2023},
  volume       = {12},
  number       = {1},
  pages        = {135},
  doi          = {10.1186/s13677-023-00383-0},
}

@article{gogouvitis2020a,
	title        = {{Seamless computing in industrial systems using container orchestration”}},
	author       = {{Gogouvitis, S.V. and Mueller, H. and Premnadh, S. and Seitz, A. and Bruegge, B.}},
	journal      = {{Future Generation Computer Systems}},
	volume       = 109,
	pages        = {678–688},
	citation-number = 45,
	year         = 2020,
	language     = {en}
}

@inproceedings{enel-a,
  author       = {Şenel, Burak Can and Mouchet, Mathieu and Cappos, Justin and Friedman, Tim and Fourmaux, Olivier and McGeer, Rick},
  title        = {{Federating EdgeNet with Fed4FIRE+ and Deploying Its Nodes Behind NATs}},
  booktitle    = {{2022 IFIP Networking Conference (IFIP Networking)}},
  publisher    = {IEEE},
  year         = {2022},
  pages        = {1--5},
  doi          = {10.23919/IFIPNetworking55402.2022.9829856},
}

@article{casalicchio2019a,
	title        = {{Container orchestration: A survey}},
	author       = {{Casalicchio, Emiliano}},
	journal      = {{Systems Modeling: Meth-odologies and Tools}},
	pages        = {221–235},
	citation-number = 48,
	year         = 2019,
	language     = {en}
}

@article{hong2019a,
  author       = {Hong, Chun-Hsien and Varghese, Blesson},
  title        = {{Resource Management in Fog/Edge Computing: A Survey on Architectures, Infrastructure, and Algorithms}},
  journal = {{ACM Computing Surveys}},
  year         = {2019},
  volume       = {52},
  number       = {5},
  pages        = {1--37},
  doi          = {10.1145/3326066},
}

@inproceedings{larsson2020b,
	title        = {{Decentralised ku-bernetes federation control plane}},
	author       = {{Larsson, Lars and Gustafsson, Harald and Klein, Cristian and Elmroth, Erik}},
	booktitle    = {2020 IEEE/ACM 13th International Conference on Utili-ty and Cloud Computing (UCC},
	publisher    = {{IEEE}},
	pages        = {354–359},
	citation-number = 54,
	year         = 2020,
	language     = {en}
}

@article{baldoni2021a,
  author       = {Baldoni, Gabriele and Cominardi, Luca and Groshev, Milan and Oliva, Antonio and Corsaro, Angelo},
  title        = {{Managing the Far-Edge: Are Today's Centralised Solutions a Good Fit?}},
  journal = {{IEEE Consumer Electronics Magazine}},
  year         = {2021},
  volume       = {12},
  number       = {3},
  pages        = {51--61},
  doi          = {10.1109/MCE.2020.3031507},
}

@article{DHT,
	title        = {{Chord: a scalable peer-to-peer lookup protocol for internet applications}},
	author       = {{Stoica, Ion and Morris, Robert and Liben-Nowell, David and Karger, David R. and Kaashoek, M.Frans and Dabek, Frank and Balakrishnan, Hari}},
	journal      = {{IEEE/ACM Trans. Netw}},
	volume       = {11, 1},
	pages        = {17–32},
	doi          = {10.1109/TNET.2002.808407},
	citation-number = 63,
	year         = 2003,
	language     = {en}
}

@inproceedings{codef2024,
	title        = {{An Open-Source Experimentation Framework for the Edge Cloud Continuum}},
	author       = {{G. Koukis, S. Skaperas, I. Angeliki Kapetanidou, V. Tsaoussidis, L. Mamatas}},
	year         = 2024,
	booktitle    = {{\textit{INFOCOM 2024 CNERT: Computer and Networking Experimental Research using Testbeds Workshop (CNERT)}}},
	pages        = {},
	url          = {https://doi.org/10.5281/zenodo.10840008},
	organization = {IEEE}
}

@article{senjab2023survey,
	title        = {{A survey of Kubernetes scheduling algorithms}},
	author       = {{Senjab, Khaldoun and Abbas, Sohail and Ahmed, Naveed and Khan, Atta ur Rehman}},
	year         = 2023,
	journal      = {{\textit{Journal of Cloud Computing}}},
	publisher    = {{Springer}},
	volume       = 12,
	number       = 1,
	note         = {Art. no. 87}
}

@techreport{CODECO-D11,
	title        = {{CODECO D11: CODECO Basic Operation and Open Toolkit v1.0}},
	author       = {{N. Psaromanolakis (Ed.) et al.}},
	year         = 2023,
	month        = oct,
	publisher    = {{Zenodo}},
	doi          = {},
	url          = {https://zenodo.org/records/10148463},
	institution  = {CODECO Deliverables Series},
	version      = {1.0}
}

@inproceedings{mueller-a,
	title        = {{Seamless Computing for Industrial Systems Spanning Cloud and Edge}},
	author       = {{Mueller, H. and Gogouvitis, S.V. and Seitz, A. and A., B.Bruegge}},
	year         = 2017,
	booktitle    = {\textit{{2017 International Conference on High Performance Computing \& Simulation (HPCS}}},
	publisher    = {{IEEE}},
	pages        = {209–216},
	citation-number = 5,
	language     = {en}
}

@article{zheng2022a,
  author       = {Zheng, Tong and Wan, Jian and Zhang, Jian and Jiang, Congfeng},
  title        = {{Deep Reinforcement Learning-Based Workload Scheduling for Edge Computing}},
  journal = {{Journal of Cloud Computing}},
  year         = {2022},
  volume       = {11},
  number       = {1},
  doi          = {10.1186/s13677-021-00276-0},
}

@article{xiong2020a,
  author       = {Xiong, Xiong and Zheng, Kan and Lei, Lei and Hou, Lu},
  title        = {{Resource Allocation Based on Deep Reinforcement Learning in IoT Edge Computing}},
  journal = {{IEEE Journal on Selected Areas in Communications}},
  year         = {2020},
  volume       = {38},
  number       = {6},
  pages        = {1133--1146},
  doi          = {10.1109/JSAC.2020.2986615},
}

@incollection{bu2010a,
  author       = {Buşoniu, Lucian and Babuška, Robert and De Schutter, Bart},
  title        = {{Multi-Agent Reinforcement Learning: An Overview}},
  booktitle    = {{Innovations in Multi-Agent Systems and Applications}},
  publisher    = {Springer},
  address      = {Berlin and Heidelberg},
  year         = {2010},
  pages        = {183--221},
  doi          = {10.1007/978-3-642-14435-6_7},
}

@incollection{zhang-a,
	title        = {{Multi-agent reinforcement learning: A selective overview of theories and algorithms”}},
	author       = {{Zhang, K. and Zhuoran, Y. and Basar, T.}},
	year         = 2021,
	booktitle    = {Handbook of reinforcement learning and control”},
	pages        = {321–384},
	citation-number = 15,
	language     = {en}
}

@online{browne2019a,
  author       = {Browne, J. and others},
  title        = {{Closed-Loop Automation: Telemetry-Aware Scheduler for Service Healing and Platform Resilience}},
  year         = {2019},
  organization = {Intel},
  url          = {https://bit.ly/3RgHmEP},
}

@inproceedings{huang-a,
	title        = {{RLSK: A Job Scheduler for Federated Kubernetes Clusters based on Reinforcement Learning}},
	author       = {{Huang, Jiaming and Xiao, Chuming and Wu, Weigang}},
	year         = 2020,
	booktitle    = {2020 IEEE International Conference on Cloud Engineering (IC2E)},
	volume       = {},
	number       = {},
	pages        = {116--123},
	doi          = {10.1109/IC2E48712.2020.00019}
}

@article{qi-b,
	title        = {{Assessing Container Network Interface Plugins: Functionality, Performance, and Scalability”}},
	author       = {{S. Qi, S.G.Kulkarni, K.K. Ramakrishnan}},
	journal      = {{IEEE Transactions on Network and Service Management}},
	volume       = 18,
	number       = 1,
    year         = 2021,
	pages        = {656–671},
	citation-number = 68,
	language     = {en}
}

@inproceedings{camaz2020a,
	title        = {{Traffic Optimization at the Application-Level Proof of concept, development and usefulness evaluation of the ALTO solution}},
	author       = {{Camaz, G. and Caldas, P. and Sousa, P.}},
	booktitle    = {15th Iberian Conference on Information Systems and Technologies (CISTI), Seville},
	address      = {Spain},
	pages        = {1–6,},
	doi          = {10.23919/CISTI49556.2020.9140947.},
	citation-number = 70,
	year         = 2020,
	language     = {en}
}

@incollection{ghiran-a,
	title        = {{The Model-Driven Enterprise Data Fabric: A Proposal Based on Conceptual Modelling and Knowledge Graphs}},
	author       = {{Ghiran, A.M. and Buchmann, R.A.}},
	year         = 2019,
	booktitle    = {\textit{{Knowledge Science, Engineering and Management. KSEM 2019. Springer Lecture Notes in Computer Science}}},
	publisher    = {{Springer}},
	address      = {Cham},
	volume       = 11775,
	doi          = {10.1007/978-3-030-29551-6_51},
	citation-number = 81,
	language     = {en}
}

@INPROCEEDINGS{pathfinder,
 author       = {Rooney, Sean and Garc{\'e}s-Erice, Luis and Bauer, Daniel and Urbanetz, Peter},
  title        = {{Pathfinder: Building the Enterprise Data Map}},
  booktitle    = {{2021 IEEE International Conference on Big Data (Big Data)}},
  publisher    = {IEEE},
  year         = {2021},
  pages        = {1909--1919},
  doi          = {10.1109/BigData52589.2021.9671608},
}

@article{bellavista2021,
	title        = {{Decentralised Learning in Federated Deployment Environments: A System-Level Survey}},
	author       = {{Bellavista, Paolo and Foschini, Luca and Mora, Alessio}},
	year         = 2021,
	month        = feb,
	journal      = {{ACM Comput. Surv.}},
	publisher    = {{Association for Computing Machinery}},
	address      = {New York, NY, USA},
	volume       = 54,
	number       = 1,
	doi          = {10.1145/3429252},
	issn         = {0360-0300},
	url          = {https://doi.org/10.1145/3429252},
	issue_year   = {January 2022},
	articleno    = 15,
	numpages     = 38
}

%

\balance
\end{document}